# All-optical intracellular thermal profiling using nanodiamond-based "thermal radar"

Jiahua Zhang[1,†], Yong Hou[1,†], Xinhao Hu[1], Yicheng Wang[1], Madoka Suzuki[3], Bo Gao[4,5,*], Zhiqin Chu[1,2,*]

[†] Equal contribution

[*] Corresponding authors (Email: bogao@cuhk.edu.hk; zqchu@eee.hku.hk)

[1] Department of Electrical and Electronic Engineering, The University of Hong Kong, Hong Kong, China

[2] School of Biomedical Sciences and School of Biomedical Engineering, The University of Hong Kong, Hong Kong, China

[3] Institute for Protein Research, The University of Osaka, Osaka, Japan

[4] School of Biomedical Sciences, Faculty of Medicine, The Chinese University of Hong Kong, Shatin, Hong Kong, China

[5] Centre for Translational Stem Cell Biology, Tai Po, Hong Kong, China

## Abstract:

The local thermal conductivity ($\kappa$) is a pivotal biophysical parameter, governing intracellular heat flux and underlying functional processes like metabolic regulation and stress response. However, label-free mapping with sub-micron resolution in living cells remains challenge. Here, we present frequency-domain fluorescence thermometry (FD-FTM), an all-optical method based on a hybrid nanodiamond–on–gold-membrane platform, which enables quantitative mapping of $\kappa$ in biological systems. Fluorescence nanodiamonds (FNDs) are deposited on substrates coated with a 50 nm gold membrane, where FNDs function as nanoscale thermometers, and the gold membrane serves as a photothermal heat source. We validate FD-FTM across reference materials and biological media, with fitting uncertainties of ~10%. By varying the modulation frequency, we tune the thermal penetration depths, enabling controlled heat propagation from the substrate to the cell nucleus. The method delivers sensitivity sufficient to resolve changes in biofluid thermal conductivity on the order of 16% relative to water. Using these capabilities, we demonstrate non-invasive thermal profiling across scales: at the cellular level, nuclear chromatin packing yields $\kappa$ higher by ~10% relative to the cytoplasm; at the organelle level, we resolve $\kappa$ variations associated with protein aggregates formed during liquid–liquid phase separation in an amyotrophic lateral sclerosis disease model. Temporal measurements in living cells over 30 minutes further reveal spatially resolved intracellular responses to osmotic stress, linking nanoscale thermal dynamics to biomolecular condensates. These results establish FD-FTM as a label-free, robust, and quantitative platform for thermally decoding intracellular processes, opening avenues for studying metabolic heterogeneity, disease mechanisms, and therapeutic responses.

## Introduction:

The functional and pathological state of a cell is governed by the complex, heterogeneous organization of its internal components [1]. Recent single-cell studies indicate that the associated thermal environment is likewise heterogeneous and strongly dependent on cellular state. Intracellular measurements of heat transport report effective thermal conductivities, $\kappa$, in living cells that are substantially lower than that of water, typically on the order of 0.1–0.4 W·m$^{-1}$·K$^{-1}$, with broad distributions that reflect pronounced variability within individual cells and between different physiological states [2,3]. Within a single cell, $\kappa$ has been found to span a wide range across subcellular regions, with systematically higher values near the plasma membrane than in more central regions, and with broader $\kappa$ distributions in living cells than in fixed counterparts, indicating that both composition and active processes contribute to the observed heterogeneity [3]. At larger length scales, macroscopic measurements likewise reveal reduced effective $\kappa$ in malignant compared with normal tissues, consistent with altered water content, macromolecular packing, and structural organization in disease [4]. Together, these observations establish that heat-transport properties in biological systems are not uniform material constants but sensitive reporters of cellular composition and state.

Currently, the dominant approach for subcellular thermal conductivity sensing in living cell is the luminescence thermometry based on nanomaterials [2,3] or polymers [5]. Within this paradigm, Sotoma et al. developed heater–thermometer hybrid diamond nanosensors for in situ intracellular measurements. The unique "two-in-one" nanoheater/nanothermometer design enabled a calculated $\kappa$ in living cells by assuming living cells are a uniform body [2]. In contrast, Song et al. used transient plasmonic imaging of the intracellular gold nanoparticles to map heat transfer and thermoregulation in single cells, uncovering anisotropic conduction that contradicts uniformity assumptions [3]. These studies mark a progression from uniform models toward recognition of pronounced thermal heterogeneity at the subcellular level, underscoring the need for methods that resolve spatial variations in $\kappa$. Despite this progress, these intracellular methods inherently require the physical delivery of exogenous nanomaterials. This might raise legitimate concerns about physiological perturbation, potential cytotoxicity, and altered cellular functions. Such issues preclude long-term, non-invasive studies and potentially confound the very measurements they aim to make. Thus, a non-invasive alternative is urgently needed to probe thermal heterogeneity in living cells without perturbation.

An ideal solution for non-invasive, depth-resolved sensing applications would integrate heating and remote temperature readout in a single platform, eliminating the need for internalized probes. This concept finds a powerful analogue in the field of materials science: frequency-domain thermoreflectance (FDTR) [6,7]. FDTR relies on resolving the minute temperature-dependent reflectance changes (~$10^{-4}$ /K) of the targeted materials [8,9]. By employing an intensity-modulated pump laser that generates a "thermal wave" within the sample, one can control the penetration depth by varying the modulation frequency [10], functioning as a precise "thermal-wave radar" for depth-resolved thermal conductivity measurement [11-13]. However, this technique relies on a high optical power (tens of mW to hundreds of mW) and multi-micrometer spot sizes [14,15], creating a seemingly insurmountable barrier to living samples.

Here, we overcome this barrier by fundamentally redesigning the conventional FDTR concept and introduce a novel all-optical frequency-domain fluorescence thermometry (FD-FTM) technique. The core innovation lies in replacing the low-sensitivity reflectance readout with fluorescent nanodiamonds (FNDs) as highly sensitive, biocompatible nanothermometers. Importantly, the temperature-dependent fluorescence signal of FNDs (~ 1% /K) [16] is two orders of magnitude larger than temperature-dependent reflectance, enabling us to achieve µW-level probe powers compatible with living cells. We validate the method against reference materials and demonstrate its capability to map intracellular $\kappa$ with subcellular specificity. The technique resolves differences between the cytoplasm and nucleus at the single-cell level and offers sufficient spatial resolution to distinguish protein concentrations during liquid–liquid phase separation in an amyotrophic lateral sclerosis (ALS) disease model. In addition, we perform time-resolved tracking of intracellular dynamic responses to osmotic stress in living cells. Compared to existing thermal conductivity measurement techniques (Table in Fig. 1), our work establishes a robust, high-resolution, non-invasive platform for quantitative thermal profiling of biological systems, with broad implications for biomedical research, diagnostics, and therapeutic development.

## Results:

**Frequency domain fluorescence-thermometry for thermal conductivity measurement**

To implement our FD-FTM method, the schematic diagram is shown in Fig. 1. We followed transient thermoreflectance methodologies and deposited a 50 nm gold membrane on the sample, which was irradiated by a continuous-wave green laser to serve as the localized heating source. FNDs with 150nm diameter were then positioned on the surface of the gold membrane and used as nanothermometers. The single green laser simultaneously heated the gold membrane and excited FND fluorescence, eliminating the need for a dual-beam pump-probe setup. This single-beam design simplifies instrumentation and enhances compatibility with commercial fluorescence confocal microscopes. The optical path is shown in Fig. S1.

When the heating laser is sinusoidally modulated, the fluorescence oscillates at the same frequency. In our setup, fluorescence was collected by an avalanche photon diode (APD) and processed via time-correlated single photon counting (TCSPC) to obtain the signal waveform. In the absence of a heat conduction layer and without laser-induced heating, the modulation amplitude remains constant across low frequencies, corresponding to the well-known frequency-domain lifetime curve (black curve in Fig. 1). With laser-induced heating, the temperature rise induces a linear decrease in fluorescence intensity for FNDs, following the relationship $I_T = I(1 - \alpha T)$ where $\alpha \approx 1\%$ /K between approximately 298 K and 328 K [16]. The amplitude of the oscillating heating temperature field decreases with increasing modulation frequency, inversely affecting the fluorescence signal amplitude. With a gold membrane, the fluorescence modulation amplitude increases at low frequency due to larger temperature excursions (blue curve in Fig. 1). Lower thermal conductivity $\kappa$ results in larger temperature rises and thus steeper frequency dependence of the fluorescence modulated amplitude (red curve in Fig. 1). By combining a thermal model with fluorescence lifetime analysis, we fit the frequency-domain fluorescence modulation curves to extract the thermal conductivity of various materials and cells. Details of the fitting theory are provided in the Supporting note 3.

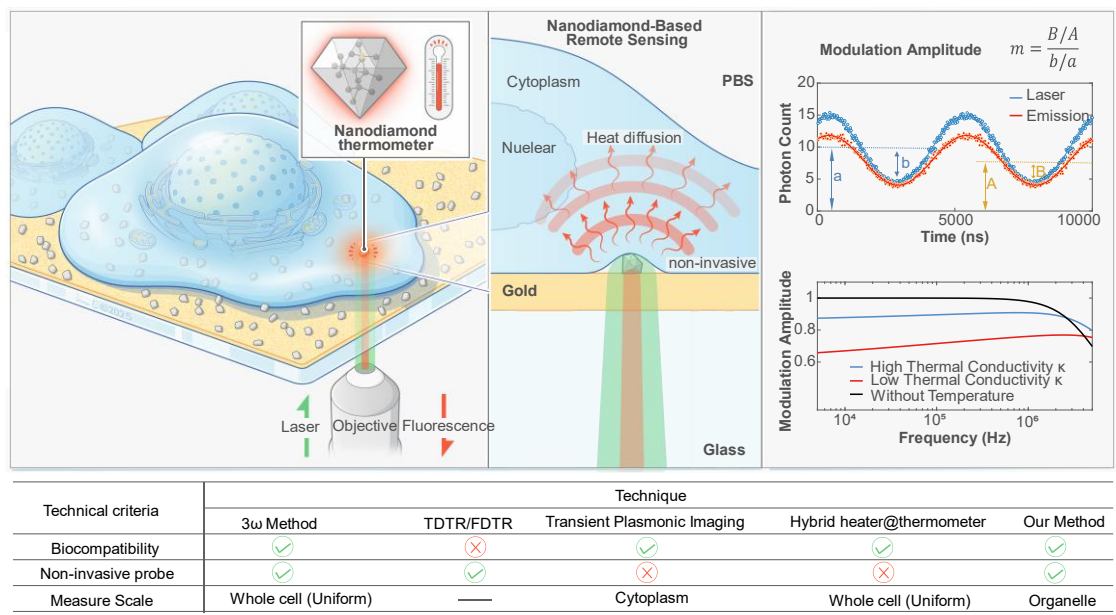

Figure 1. **Conceptual illustration of frequency domain fluorescence thermometry.** The sinusoidally modulated laser heats the gold membrane, from which the heat diffuses bidirectionally into the overlying cells and the underlying gold-membrane/glass-substrate stack. In the illustration, we depict only the heat flux into the cell to emphasize the scheme's intracellular remote-sensing capability. At the same time, the laser excites the fluorescence of nanodiamonds, transmitting the thermal information within fluorescence waveform. The right panel shows one frequency scenarios: the laser (blue line) is modulated sinusoidally at 5 kHz, and the resulting fluorescence (red line) oscillates at the same frequency. The response of modulation amplitudes over a broad frequency range 5 kHz – 5 MHz differs between scenarios with and without the gold membrane, where thermal conductivity can be decoded from the fluorescence modulation amplitude. In the table, our technique is compared to other existing biomedical thermal conductivity measurement techniques, offering a biocompatible, non-invasive platform for organelle level biomedical research.

## Characteristics of measurement methodology

We validated the methodology by measuring the thermal conductivity of well-characterized media: a 150 μm thick cover glass slide, a 1mm thick sapphire ($Al_2O_3$) wafer, a 500 μm thick silicon wafer, and deionized (DI) water. For all experiments, including the subsequent cellular measurements, the laser power was sinusoidally modulated with an average value of 119 μW and a modulation ratio of ~0.54. The beam was focused to a diffraction-limited spot, with a radius of approximately 400 nm in water and in the cellular experiments (NA=0.8), and 350 nm in solid samples (NA=0.95). Under these conditions, the corresponding power density was approximately 0.24 MW/m², which is about 10 times lower than that used in frequency-domain thermoreflectance (approximately 2 MW/m² [17,18]). The frequency-dependent laser modulation amplitude curve is provided in Fig. S2b. This power level was selected to remain below the saturation power of the Nitrogen-Vacancy (NV) color center (approximately 350 μW; Fig. S3a), avoiding nonlinear effects on fluorescence amplitude due to saturation. Bulk solid samples coated with a 50 nm gold membrane were characterized using our developed method, and the resulting modulation amplitude curves were fitted with a multilayer thermal model (the experiment illustration of layer structures are shown in Fig. S4a). In contrast, for the liquid measurements, 4 mL of deionized water was used to directly immerse a cover glass coated with a gold membrane of the same thickness, and the corresponding curves were fitted using a bidirectional heat conduction model. Because the heating laser beam diameter was ~0.6 μm, radial heat transfer dominated thermal dissipation. Consequently, the measured thermal conductivity corresponds to the geometric average thermal conductivity $\sqrt{\kappa_r \kappa_z}$ for anisotropic materials [19,20]. The measured thermal conductivities are listed in Table 1. Standard deviations were calculated via the variance-covariance matrix (See Methods) [21]. These results are compared against literature data, spanning over three orders of magnitude, showing excellent agreement, as shown in Fig. 2a.

It should be noted that NV-based nanothermometers are less suitable for materials with high thermal conductivity ($\kappa > 200$ W·m$^{-1}$·K$^{-1}$). Achieving comparable temperature rises in such media requires substantially higher heating powers, which can exceed the fluorescence saturation power of NV centers and introduce nonlinearities that bias the measured modulation amplitude. By contrast, cells are generally considered low-$\kappa$ media and thus fall within the optimal operating regime of NV-based nanothermometry. For high-$\kappa$ targets, nanothermometers with higher saturation powers, such as SiV centers in diamond (saturation power ~5 mW) [22], are preferable, because they preserve linear fluorescence response and improve measurement accuracy.

Table 1. Fitted thermal conductivity for samples tested using frequency-domain fluorescence-thermometry technique. The fitted $\kappa$ corresponds to $\kappa \equiv \sqrt{\kappa_r \kappa_z}$.

| Sample | $\kappa$ (W·m$^{-1}$·K$^{-1}$) | |
|---|---|---|
| | Measured | Literature |
| water | 0.59 ± 0.08 | 0.6, Ref 23 |
| Cover glass slide | 1.34 ± 0.14 | 1.3, Ref 24 |
| Al$_2$O$_3$ | 30.74 ± 2.17 | 29.5, Ref 25 |
| Silicon wafer | 138.10 ± 37.40 | 133, Ref 25 |

Following validation on well-characterized solids and liquids, we systematically quantified the uncertainty and precision of FD-FTM method to establish performance bounds for biological measurements. The analysis addressed two factors: (i) fitting precision for single shot measurement and (ii) heterogeneity among fluorescent nanodiamonds (FNDs) for statistical analyses. (i) Fitting precision is governed by uncertainty in the modulation amplitude $m_\omega$, which scales with integration time $t$ as $\sigma_m \propto 1/\sqrt{t}$. At 5 kHz for a single FND with ~ 1 million photocounts/s, the standard deviation of modulation amplitude is around 0.2% for 120 s integration per frequency, yielding smaller than 10% precision in thermal conductivity fits across 16 frequencies spanning 5 kHz - 5 MHz (Fig. S2c). Frequency sampling and the maximum frequency were limited by the signal generator (Supporting Note 1). Enhancing the FND fluorescence intensity can reduce per-frequency integration time and accelerates measurements. (ii) In addition to fitting precision, the thermal response coefficient is affected by heterogeneity among FNDs, which in turn limits measurement precision. We therefore quantified the impact of FND heterogeneity via statistical analysis by measuring deionized (DI) water with 25 individual FNDs on the same sample (Fig. 2b). The heterogeneity-induced precision (~11%, 0.068 W·m$^{-1}$·K$^{-1}$) was comparable to the fitting precision (~13%, 0.080 W·m$^{-1}$·K$^{-1}$), indicating minimal impact. For subsequent cellular measurements, we thus prioritized statistical analyses across multiple FNDs.

To further validate FD-FTM's capability to quantify key constituents of the cellular microenvironment (~18% proteins, 5% lipids, 1.35% nucleic acids and 1% salts [26]), we prepared protein and salt model systems. Specifically, we used bovine serum albumin (BSA) and phosphate-buffered saline (PBS) solutions at varying concentrations to represent proteins and inorganic salts, respectively. For each concentration, we collected data from 10 in situ FNDs. For PBS (Fig. 2c, the measured thermal conductivity of ~0.58 W·m$^{-1}$·K$^{-1}$ tends to be slightly lower than that of water (0.6 W·m$^{-1}$·K$^{-1}$), with no significant concentration dependence (1X - 4X). Although precise $\kappa$ values for each PBS concentration at room temperature are not directly reported, studies of analogous NaCl solutions (the primary PBS solute) show a modest, roughly linear decrease in $\kappa$ with increasing concentrations, typically on the order of ~ 1% (~ 0.006 W·m$^{-1}$·K$^{-1}$) between ~0.137 mol/L to ~0.548 mol/L NaCl (approximately 1X to 4X PBS) [27,28]. These expected changes lie below our setup's sensitivity and are consistent with the negligible concentration dependence observed. For BSA solution, as shown in the Fig. 2d, $\kappa$ decreases linearly from ~ 0.574 W·m$^{-1}$·K$^{-1}$ to 0.540 W·m$^{-1}$·K$^{-1}$ with increasing BSA concentration, consistent with the Maxwell-Eucken model [29]. Overall, our method reliably resolves changes in thermal conductivity on the order of 16% relative to that of water (~0.01 W·m$^{-1}$·K$^{-1}$), making it sensitive to macromolecules variations while remaining relatively insensitive to ionic shifts (~1% across 1X - 4X PBS). Together, these results support thermal sensing as a practical approach for monitoring macromolecular changes in cellular environments.

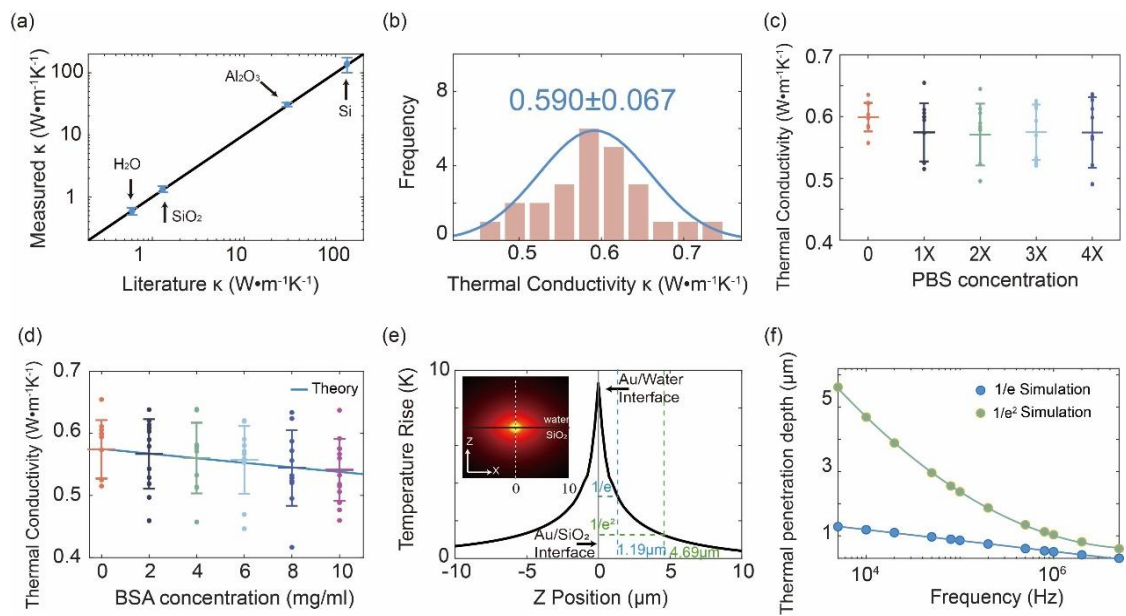

Figure 2. **Characteristics of frequency domain fluorescence thermometry.** (a) Accuracy of thermal conductivity measurement verified by four known sample. (b) Impact from FND heterogeneity. Statistical analysis of 25 FNDs fitted to a normal distribution. (c) Thermal conductivity for PBS concentration. (d) Thermal conductivity for BSA protein concentration. Data from 10-14 FNDs per condition. (e) A temperature profile under 10 kHz modulated heating (0.1mW absorbed power). The heating laser beam was set to 0.61λ/NA (NA = 0.8, λ = 0.532μm); The insert image is the temperature map for SiO$_2$/50nm Au/water under unmodulated laser. (f) Thermal penetration depth (TPD) across the measurement frequency range. 1/e TPD is standard for bulk/thin film.

Having established the reliability of our measurement methodology, we next examined how modulation frequency governs the sensing depth reach via the thermal penetration depth (TPD). TPD is critical because it relates the measurement length scale to the physical phenomena of interest, thereby setting the accessible cellular depth scales and enables remote, low-phototoxic sensing from the Au-coated interface. Because our geometry produces predominantly radial thermal diffusion, the standard in-plane TPD expression $d_f = \sqrt{\kappa / \pi f C}$, where $f$ is modulation frequency, $\kappa$ and $C$ are the thermal conductivity and volumetric heat capacity of the sample, does not apply. We therefore quantified TPD via transient finite element simulation (model and parameters in Supporting Note 4). As shown in Fig. 2e-f, TPD decreases logarithmically with frequency: the 1/e TPD falls from 1.3 μm at 5 kHz to 0.3 μm at 5 MHz, and the $1/e^2$ TPD (closer to the heating boundary) falls from 5.6 μm to 0.6 μm. These depth ranges span typical cellular dimensions, enabling remote interrogation with minimal phototoxicity and supporting thermal-wave, radar-like sensing of intracellular conditions.

For subsequent cellular measurements, we further evaluated photo- and thermal toxicity of the FD-FTM platform. We quantified the correlations of heating laser power and the resulting temperature rise caused in gold membrane using both numerical simulations and experimental validations (Fig. S6). In DI water, a 100 μW heating laser produced a 3.89 K rise at NV centers in FNDs, about half of the 6.96 K predicted by a simplified model that neglected the gold membrane's finite absorption. We then used an improved model that incorporates gold absorption and adds a homogeneous cell model to estimate temperature rises under the actual live-cell imaging conditions (Supporting Note 4). For a representative incident power of 120 μW, the simulations predicted mean steady-state temperature increases of approximately 4.69 K and 5.52 K beneath the nuclear and cytoplasmic regions, respectively, with the sinusoidal power modulation adding ~2 K oscillation about these means. The magnitude of these nanoscale temperature excursions is consistent with prior reports [5,30] and lower than those typically employed in photothermal cellular stimulation studies [31,32], suggesting that the imposed heating at 306 K, the condition used in the following live-cell experiments, is unlikely to appreciably perturb cell physiology. Consistent with the cytotoxicity data showing no significant impact on cell viability and proliferation (Fig. S7), these results demonstrate the excellent biocompatibility of our technique.

**Thermal conductivity mapping within a single cell**

With validated sensitivity, accuracy and biocompatibility, we next sought to evaluate its capability for depth-thermal profiling within cells. A key advantage of our approach is the ability to probe intracellular regions from the sensor interface, enabling non-invasive $\kappa$ mapping across cellular compartments.

We mapped heat conduction in a single human cervical cancer (HeLa) cell. HeLa cells, widely used in various biology studies, were cultured on gold membrane with FNDs attached at the surface randomly. FNDs were positioned randomly either adjacent to the basal membrane (probing the cytoplasmic region) or beneath the nuclear region, which lies several micrometers above the substrate (Fig. S8, FNDs to nuclear bottom: ~1.5 μm). Measurements were performed at the room temperature on a custom-built confocal setup. For the fixed HeLa cell, the thermal conductivity at a single spot in the scanning image was determined from a measurement lasting approximately 10 minutes. In-situ three-dimensional (3D) fluorescence scans were acquired to obtain the cell height profile, which was then used as conditions in the model fitting. Since the characteristic dimensions of the nucleus and cytoplasm far more exceed the heating spot (~0.6 μm diameter on the order of the laser wavelength), the cell can be treated as a layer medium and analysed using a bidirectional multilayer heat conduction model [33]. The measured thermal conductivity represents a volume-averaged value over the probed region, which can be approximated as the product of the lateral spatial resolution and the thermal penetration depth. Consequently, for FNDs located beneath the nuclear region, the measured thermal conductivity corresponds to an effective value that reflects contributions from both the cytosol and the nucleus. At present, the spatial resolution is limited by optical diffraction (approximately 400 nm), but it could, in principle, be reduced to the scale of the FND diameter by integrating super-resolution techniques. Fig. 3a shows the thermal conductivity map for a single HeLa cell, with $\kappa$ ranging from 0.163 to 0.366 W·m$^{-1}$·K$^{-1}$, consistent with the previous fixed result [3]. Spatial heterogeneity in $\kappa$ is evident across the cell, reflecting local variations in biomolecular concentration.

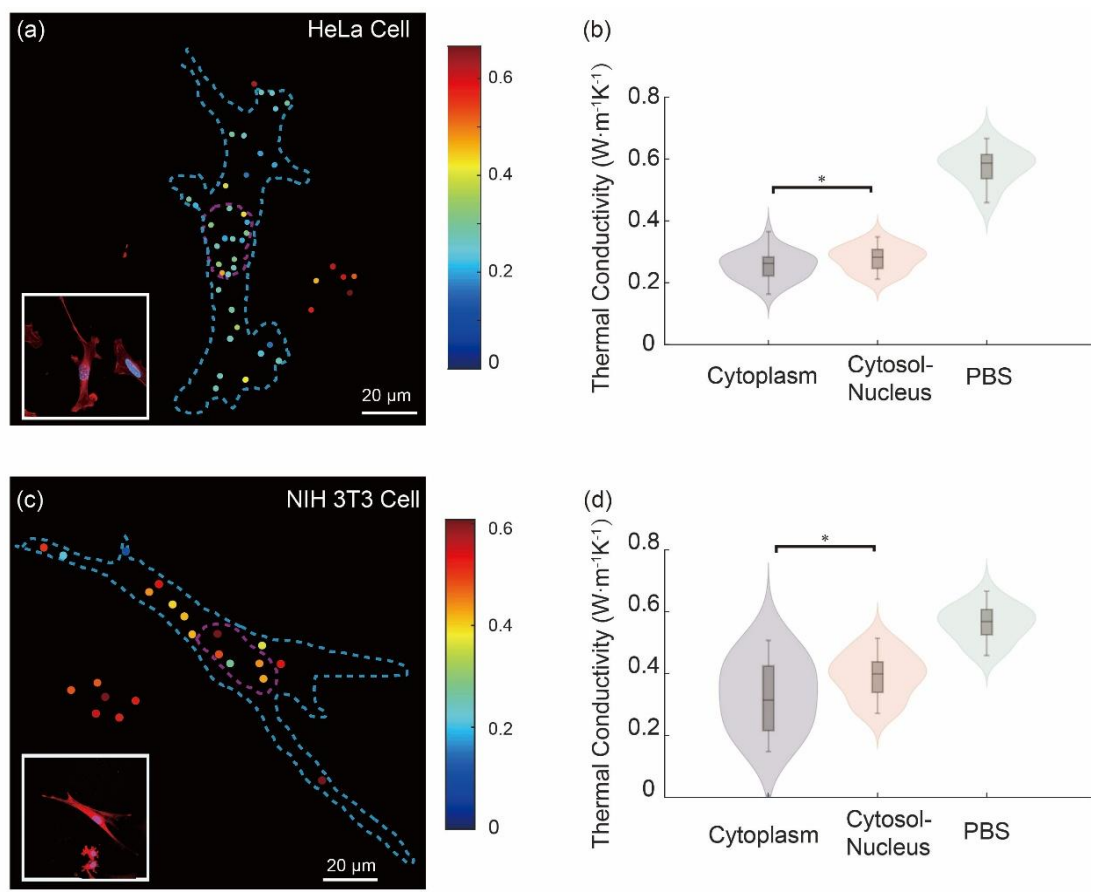

Figure 3. **Intracellular spatially distributed thermal conductivity mapping at cellular level**. Thermal conductivity map of a single fixed (a) HeLa cell and (c) NIH 3T3 cell collected from FNDs. The insert image is the corresponding confocal xy image of a fixed cell. Violin plots of thermal conductivity $\kappa$ in cytoplasm and nucleus of fixed (b) HeLa cells and (d) NIH 3T3 cells; PBS as control. The statistical number of FNDs with n=20-30 for PBS, cytoplasm and nucleus, respectively. (* for p < 0.05, p=0.0257 for HeLa cells and p=0.0269 for NIH 3T3 cells).

Critically, we first found that the region comprising the nucleus and the underlying cytoplasm, hereafter referred to as the cytosol-nuclear region, showed a significantly higher thermal conductivity (0.278 ± 0.040 W·m$^{-1}$·K$^{-1}$) than that of the cytoplasm (0.254 ± 0.045 W·m$^{-1}$·K$^{-1}$) (Fig. 3b). Using the Maxwell–Eucken model to interpret these measurements (see Methods), we estimate that the effective concentration of dry matter in the cytosol-nuclear region is approximately 4.9–6.6% lower than in the cytoplasm. This observation aligns with reports of higher water content (by 2-3%) and lower refractive index (RI) (i.e., lower dry mass density) in the nucleus relative to the cytoplasm, as supported by Raman microscopy [34,35] and interferometry studies [36-38]. In particular, refractive index measurements in HeLa cells indicate that the nucleoplasm is 5.7–7.3% less dense than the cytoplasm [37], closely matching our Maxwell–Eucken-based estimates. Thus, higher water content and lower dry matter concentrations in the nucleus would be expected to yield higher $\kappa$, corroborating our findings. To further strengthen our finding, other cell types (e.g. NIH 3T3 cell) have also been tested and the similar results were observed (Fig. 3c–d), suggesting that elevated nuclear thermal conductivity is a general feature rather than a peculiarity of HeLa cells. These results demonstrate the capacity of FD-FTM to resolve depth-dependent thermal properties in cells, validating its use as a non-invasive, spatially resolved thermal profiling tool.

**Thermal conductivity mapping in the nucleus during liquid-liquid phase separation**

Based on the capability to resolve subcellular thermal heterogeneity, we next aimed to push spatial resolution of FD-FTM to the organelle scale, targeting biomolecules aggregation formed via liquid-liquid phase separation (LLPS). LLPS drives the formation of membrane-less organelles and protein aggregates, which are implicated in numerous cellular functions and disease states [39]. Because thermal conductivity is sensitive to local biomolecular concentration, we hypothesized that $\kappa$ mapping could provide a label-free readout of LLPS dynamics within subcellular compartments.

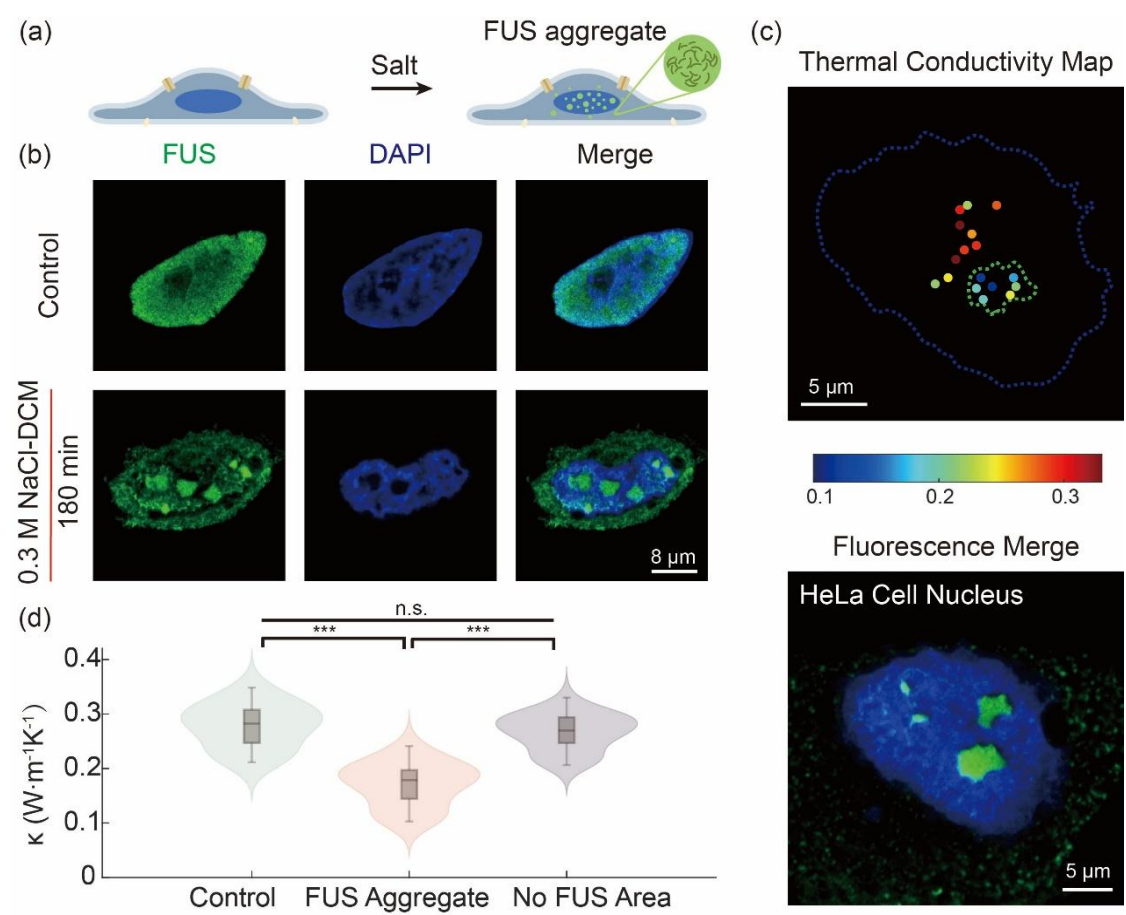

Figure 4. **Thermal conductivity mapping in the nucleus during liquid-liquid phase separation** (a) Schematic of protein fused in sarcoma (FUS) aggregation under hyperosmotic stress. (b) Confocal fluorescence image of FUS aggregation by 0.3 M NaCl dissolved in the PBS for 180 min. The fluorescence image of FUS in the complete medium without NaCl as control experiment. (c) Confocal xy fluorescence image and thermal conductivity map of a fixed HeLa cell (180 min, 0.3 M NaCl). (d) Statistical measurement of thermal conductivity for FUS aggregates and nuclear regions without FUS aggregation. The statistical number of FNDs with n=30 for control, FUS aggregates and nuclear regions without FUS aggregation, respectively. (*** for $p < 0.001$, n.s. for $p > 0.05$).

We focused on Amyotrophic Lateral Sclerosis (ALS), a neurodegenerative disease characterized by pathological LLPS of the protein fused in sarcoma (FUS) in cells [40-42]. FUS is an RNA binding protein involved in processes such as DNA repair, RNA processing, and cellular stress responses. Mutations in FUS are a known cause of familial ALS, leading to aberrant cytoplasmic mislocalization and pathological phase separation of the protein. HeLa cells were chosen as the cell model and treated with NaCl dissolved in culture medium to induce hyperosmotic stress [43]. Fig. 4 and Fig. S9a show the phase separation behaviour of FUS protein in the nucleus under hyperosmotic stress with different conditions. Under normal conditions, FUS is primarily localized in the nucleus, where it is uniformly distributed. In response to hyperosmotic stress, the FUS loses its normal distribution and assembles into granules in the nucleus and can translocate from the nucleus to the cytoplasm. The size of protein aggregation shows a NaCl concentration dependent manner. At low concentration of 0.1 M NaCl, protein granules were predominantly around 1 μm or smaller, while under 0.3 M NaCl for 180 min, the aggregates size exceed 2 μm (Fig. 4b). To ensure reliable colocalization between the FND sensor and FUS condensates within the diffraction limit of our system, we focused on nuclear FUS aggregation with size larger than 2 μm. We mapped thermal conductivity in a single fixed HeLa cell exhibiting LLPS behaviour in PBS at room temperature, and additionally performed point measurements at selected locations in the scan, with each measurement lasting approximately 10 minutes. Our results revealed that region with FUS aggregation exhibited lower $\kappa$ values (0.174 ± 0.040 $W·m^{-1}·K^{-1}$) than that on regions devoid of FUS aggregates (0.278 ± 0.040 $W·m^{-1}·K^{-1}$) (Fig. 4c, d and Fig. S9c). Nuclei without NaCl treatment exhibited a comparable $\kappa$ of 0.278 ± 0.040 $W·m^{-1}·K^{-1}$ to the non-aggregate regions in stressed nuclei, indicating that the $\kappa$ reduction is specifically attributable to the FUS condensates. The pronounced $\kappa$ contrast between FUS-aggregated and non-aggregated nuclear regions directly demonstrates that protein condensation during LLPS creates localized domains of reduced thermal conductivity, consistent with the order of 0.1–0.2 $W·m^{-1}·K^{-1}$ for proteins [44,45]. These data underscore the sensitivity of FD-FTM to microscale intracellular compositional changes in label free manner.

**Measurement of thermal conductivity in living cells for spatial and temporal monitoring**

To comprehensively assess the spatiotemporal performance of FD-FTM, we investigated its performance in dynamic live-cell imaging under osmotic stress, a well-established model that induces rapid changes in protein concentration and cellular volume. Using this perturbation, we quantified temporal fluctuations in intracellular thermal conductivity with a temporal resolution of 2 min, thus demonstrating the capability of FD-FTM for label-free and non-invasive monitoring of intracellular biophysical dynamics.

We first verified the cellular responses by applying hypo- and hyper-tonic stimuli (water and PEG 300, respectively) at varying volume fractions [46]. Cellular cytoplasm and nucleus were labelled with Calcein AM and Hoechst 33342 to monitor volume changes via confocal microscopy. Three-dimensional fluorescent images were acquired at 1 - 1.5 min intervals. As expected, hypotonic stress induced cell swelling immediately (within 1 min) followed by gradual recovery, whereas hypertonic stress caused rapid shrinkage of both the nucleus and cytoplasm (within 1 min) followed by relatively slow recovery (Fig. 5a, b, d and Fig. S10a, b, d). To ensure cell viability while achieving measurable responses, we selected mild, non-toxic conditions for subsequent FD-FTM tracking: 10% water for hypotonic stress and 5% PEG 300 for hypertonic stress for ~ 20 min live cell experiment. We tracked volumetric and thermal responses simultaneously, allowing a direct comparison between volumetric changes and the thermal conductivity dynamics. Each thermal conductivity acquisition lasted 1.5 minutes, followed by a 30 seconds interval for FND tracking and mechanical drift calibration before the next measurement.

HeLa cells were initially selected as the model system for osmotic-stress experiments. However, under hypotonic conditions, cell swelling increased the distance between the nucleus and the substrate beyond the TPD, rendering the nuclear region inaccessible to FD-FTM (as described below and Supporting Note 6 for details). Therefore, NIH 3T3 cells were used as the model for hypotonic stress, whereas HeLa cells were retained as the model for hypertonic stress. Under hypotonic stress induced by water addition in NIH 3T3 cells, cytoplasmic $\kappa$ increased rapidly, concomitant with cell swelling during the first 1-2 min, and then gradually recovered over the subsequent ~ 20 min (Fig. 5a-c). These $\kappa$ dynamics were reproducible across cells (Fig. 5c). Conversely, in HeLa cells under hypertonic stress (PEG 300 addition), nuclear $\kappa$ experienced a fast decrease synchronously with nuclear shrinkage within ~ 3 min, followed by faster recovery than under hypotonic stress. A similar trend was observed in the cytoplasm, demonstrating the same $\kappa$–volume relationship as in NIH 3T3 cells and indicating that this relationship is not cell-line specific (Fig. S10c-e).

Because single cells often migrate on the culture dish during live-cell imaging, we quantified the influence of motion on thermal conductivity tracking. During live-cell imaging, we observed that cells migrated ~1 μm over 40 minutes (small relative to cell size), resulting in only 1.5–2% signal oscillation. Thus, migration had negligible impact on dynamic thermal conductivity measurements within our observation window. However, pronounced nuclear motion under hypotonic stress (Fig. S11) prevented reliable single-point tracking of nuclear $\kappa$. Thus, time-resolved nuclear $\kappa$ was not pursued under this condition. Additional details on cell motion are provided in Supporting Note 6.

Assuming that the total amount of cellular protein remains constant, the mean relative volume change in NIH 3T3 cells implies that the protein concentration initially increases by ~14% and then partially recovers to ~6% above its baseline value. Independently, using the relative changes in thermal conductivity together with the Maxwell–Eucken effective-medium model, we estimate from our FD-FTM measurements that the intracellular biomolecular concentration increases by ~12% and then recovers to ~1% above baseline in cell 2, and from ~4% to ~2.6–2.9% above baseline in cells 1 and 3. Applying the same analysis to HeLa cell nuclei, the nuclear volume changes suggest that the protein concentration initially decreases by ~17% and then recovers to ~9% above its initial value. From the nuclear thermal-conductivity changes and the Maxwell–Eucken model, we infer that the biomolecular concentration decreases by ~21% and then recovers to ~18% below baseline in cell 1, decreases by ~28% and then overshoots to ~9% above baseline in cell 2, and decreases by ~14% before recovering to ~6% above baseline in cell 3. Given the high mobility of intracellular biomolecules, such micro- to nanoscale heterogeneity for two type cells in local concentration is plausible.

Taken together, $\kappa$ increased with volume expansion and decreased with volume shrinkage, strongly suggesting that lower biomolecular concentration (proteins and nucleic acids) leads to a higher $\kappa$, and vice versa. This correlation confirms that FD-FTM can serve as a sensitive, non-invasive method to track intracellular biophysical dynamics in living cells.

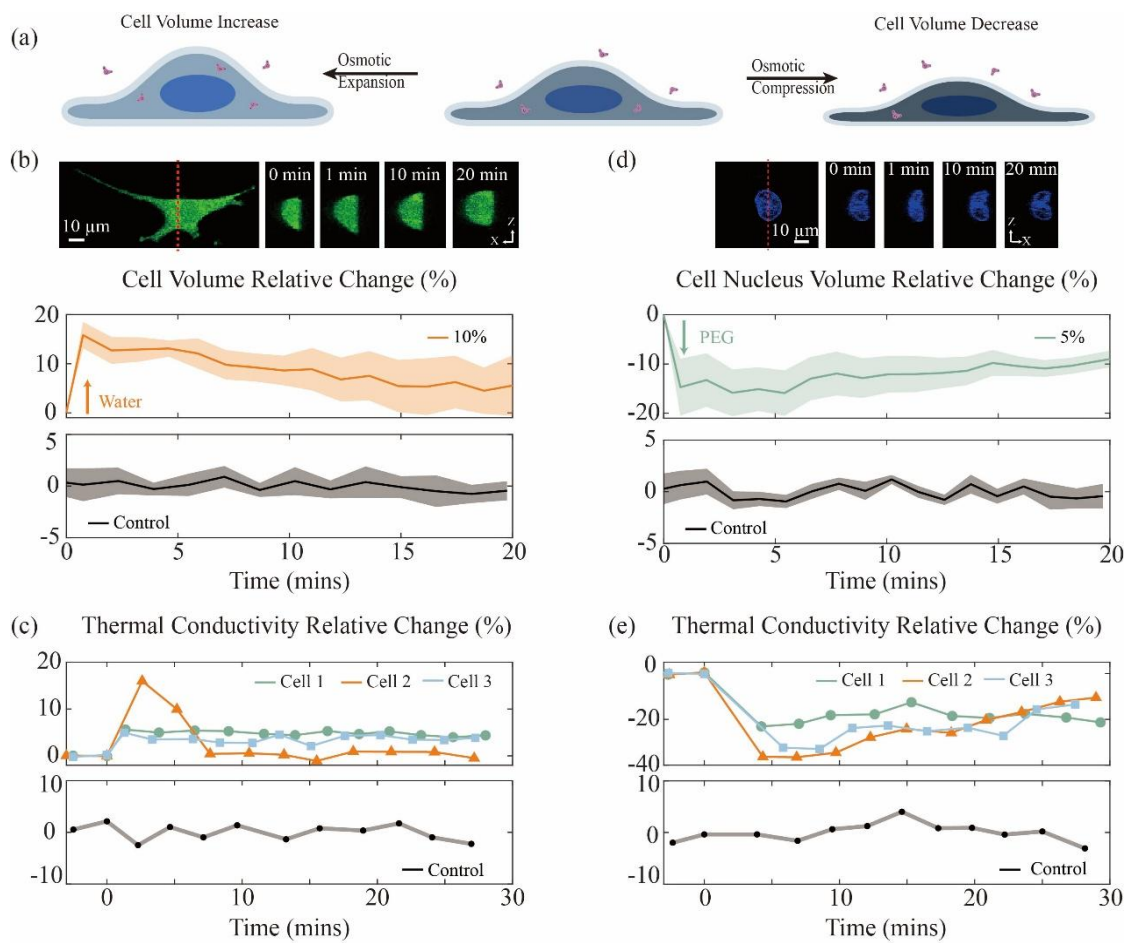

Figure 5. **Measurement of the change of thermal conductivity following osmotic stress in a single living cell.** (a) The scheme of cells under osmotic expansion, in which the cell volume increases, and under osmotic compression, in which the cell volume decreases. (b) The cell volume relative change curve of NIH 3T3 cells after adding water solutions with different volume fractions of either 0 (Control) or 10%. Four fluorescence orthogonal projections of one of the NIH 3T3 cells after adding water solution with 10% volume fractions at 0 min, 1 min, 10 min, and 20 min are listed above the curve. (c) The measurement of thermal conductivity relative change after adding water solution with 10% volume fractions. (d) The cell nucleus volume relative change curve of HeLa cells after adding PEG solutions with different volume fractions of either 0 (Control) or 5%. Four fluorescence orthogonal projections of one of the HeLa cells after adding PEG solution with 5% volume fractions at 0 min, 1 min, 10 min, and 20 min are listed above the nucleus curve. (e) The measurement of thermal conductivity relative change after adding PEG solution with 5% volume fractions. Thermal conductivity measurements were performed in L-15 medium at 33 °C.

## Discussion:

Our work establishes frequency domain fluorescence thermometry (FD-FTM) as a versatile platform for quantitative thermal profiling of living cells. This technology employs thermally generated heat diffusion, like a "thermal wave radar", to non-invasively interrogate the intracellular space, and translate these local heat transport properties into a novel, label-free contrast mechanism for probing intracellular biophysical states. This allows us to effectively turn thermal conductivity ($\kappa$) into a quantitative readout for local biomolecular concentration and packing density, opening a previously inaccessible dimension for cellular analysis.

The non-invasive nature of our thermal-wave radar stems from a fundamental re-engineering of the measurement paradigm, shifting from intrusive probing to remote sensing. By synergizing the depth-profiling principle of FDTR with the high-sensitivity thermometry of NV centers in FNDs, we have overcome the limitations of previous intracellular thermometry methods. Our approach remotely generates and senses "thermal waves" from an external, cell-compatible substrate (the FNDs@Au-membrane sensor). This unique remote-design enables depth-profiling interrogation of intracellular structures up to several micrometers deep (e.g., ~1-5 μm for the nucleus), completely bypassing the need for physical intrusion and thus eliminating associated cytotoxicity and perturbation. In addition, the fixed-thermometer design minimizes measurement artifacts from particle motion and rotation within the heterogeneous intracellular environment, thereby enhancing measurement accuracy. This paradigm was realized by the choice of the NV center in FNDs as the thermometers. Unlike other nanothermometers (e.g., organic dyes prone to photobleaching, or quantum dots with potential heavy-metal toxicity), NV centers offer a unique photostability, inherent biocompatibility, and a high temperature sensitivity of ~1% /K. This sensitivity is two orders of magnitude greater than the reflectance change utilized in conventional thermoreflectance, allowing us to replace the high-power densities (~ 2 MW/m²) typically required in those methods with ultra-low power densities (~ 0.2 MW/m²), further ensuring biocompatibility for long-term live-cell studies.

This capability allows us to directly link intracellular thermal properties to fundamental cellular organization and state. We consistently observed a higher $\kappa$ in the cytosol-nucleus compared to the cytoplasm across two cell types. This finding provides the first label-free, quantitative thermal evidence supporting the model of the nucleus as a highly hydrated, less densely packed compartment. This distinct physicochemical environment may be suitable for fundamental nuclear functions, including chromatin dynamics and nucleic acid transactions [47-50]. Furthermore, we demonstrate that $\kappa$ serves as a sensitive, intrinsic reporter for biomolecular condensation driven by liquid-liquid phase separation. The significant reduction in $\kappa$ within the region of cytosol-FUS protein aggregates shows that FD-FTM can not only identify condensates but also quantify their altered physical state. The lower $\kappa$ directly reflects increased biomolecular packing density. This capability is particularly powerful because it operates without fluorescent labels, which can themselves influence LLPS dynamics. Consequently, our approach provides a pristine, quantitative lens to study the biophysics of condensates. Crucially, these findings raise several compelling questions: How can this ability to measure the physical properties (such as $\kappa$) of protein condensates help us distinguish functional, liquid-like condensates from pathological, solid-like aggregates? Could tracking the temporal evolution of $\kappa$ provide a label-free signature of the liquid-to-solid transition implicated in neurodegenerative disease progression? And ultimately, how might this inform the development of therapies aimed at modulating these physical states in diseases like ALS?

While precise and robust across scales, our method faces three practical limitations: TPD, temporal resolution, and spatial information density. TPD is constrained by the modulation frequency and heating spot size; in our single-beam design (shared heating and fluorescence excitation), enlarging the heating spot to increase TPD reduces fluorescence collection efficiency. Decoupling heating and excitation into a pump–probe configuration with independently optimized spot sizes would allow lower modulation frequencies and larger heating spots for deeper probing while preserving collection efficiency, extending applicability to thicker tissues and potentially whole organisms. Temporal resolution is limited by TCSPC waveform acquisition with software-based amplitude extraction, yielding in single thermal conductivity point being updated only every ~1–2 minutes. The measurement accuracy of $\kappa$ depends on the number of modulation frequencies used: increasing the number of frequency points generally improves the robustness of the thermal conductivity fit. However, while maintaining a required level of fitting accuracy, the number of frequency points can be judiciously reduced (while still exceeding the number of unknown fitting parameters), thereby shortening the acquisition time for each $\kappa$ measurement. In the future, hardware demodulation (for example, lock-in detection) could reduce integration times to the millisecond–microsecond range; paired with galvanometer-based scanning, this would enable real-time, high-throughput imaging over large fields of view. The attainable spatial information density is constrained by the stochastic distribution of FNDs, leading to incomplete coverage with interspersed gaps. Integration with super-resolution modalities would improve localization and resolve smaller, heterogeneous structures, while deploying bulk diamond substrates [51], nanodiamond films [52], or patterned nanodiamond arrays [53,54] would provide uniform, high-density sampling for comprehensive mapping. Collectively, these upgrades would overcome current TPD, timing, and coverage constraints, enabling high-density, rapid thermal conductivity imaging without sacrificing non-invasiveness or sensitivity.

From a broader perspective, the implications of this technology extend across fundamental biology and translational medicine. The ability to map "thermal fingerprints" non-invasively could revolutionize our understanding of cellular metabolic heterogeneity, particularly in cancer and development. Moreover, the capacity to monitor local biomolecular concentration in real time provides a powerful tool for drug discovery and disease mechanism studies, enabling the label-free assessment of therapeutic effects on pathological protein aggregation in neurodegenerative diseases. We envision that the accumulation of such data will catalyze the creation of a "thermal property database" for cells and tissues, providing a new biophysical metric for health and disease. By providing a unique, label-free lens to observe the intricate thermal landscape of the living cell, FD-FTM offers a transformative paradigm for biological discovery and holds profound promise for guiding future biomedical innovation.

## Methods

### Uncertainty analysis

We employed least squares regression to fit experiment data to the theoretical model. The standard deviation of measured thermal conductivities was calculated via the variance-covariance matrix [21]:

$$\text{Var}[\hat{\kappa}_U] = \left(J_U^{*'}J_U^*\right)^{-1} J_U^{*'} \left(\text{Var}[m_{\omega,T}] + J_C^* \text{Var}[X_c] J_C^{*'}\right) J_U^* \left(J_U^{*'}J_U^*\right)^{-1}$$

Where $J_U^*$ is the Jacobian matrix of uncertainty parameters, $J_C^*$ is the Jacobian matrix of certainty parameters, $\text{Var}[m_{\omega,T}]$ is the variances matrix of the modulation amplitude of fluorescence signal, and $\text{Var}[X_c]$ is the variances matrix of certainty parameters.

### BSA solution analysis

We employed the Maxwell-Eucken model [29] to compare the experiment results:

$$\kappa_{eff} = \kappa_{PBS} \frac{2\kappa_{PBS} + \kappa_{BSA} - 2\phi(\kappa_{PBS} - \kappa_{BSA})}{2\kappa_{PBS} + \kappa_{BSA} + \phi(\kappa_{PBS} - \kappa_{BSA})}$$

Where $\phi$ is volume fraction of the BSA protein, $\kappa_{BSA} = 0.267\ \text{Wm}^{-1}\text{K}^{-1}$ [55], and $\kappa_{PBS} = 0.58\ \text{Wm}^{-1}\text{K}^{-1}$ from our measurement.

The dry matter (e.g., protein and nucleic acid) within the cell was treated as an equivalent homogeneous substance, and the Maxwell–Eucken model was employed to estimate changes in its concentration based on the experimental thermal conductivity data (Fig. 3b). Assuming a thermal conductivity of $0.1 - 0.2\ \text{W·m}^{-1}\text{·K}^{-1}$ for this substance, the concentration in the nuclear region is $4.9 - 6.6\ \%$ lower than in the cytoplasm. These values are consistent with independent estimates from refractive index measurements, which indicate a $5.7 - 7.3\ \%$ lower concentration in the nucleus compared with the cytoplasm in HeLa cells.

## Cell culture

HeLa and NIH 3T3 cells were cultured in Dulbecco's Modified Eagle Medium (DMEM) supplemented with 10% (v/v) fetal bovine serum (FBS) and 1% penicillin-streptomycin at 37°C in a humidified atmosphere containing 5% $CO_2$. During live-cell imaging experiments, cells were seeded in confocal dishes and cultured in a living cell supporting platform.

## Osmotic treatment

All osmotic solutions were prepared under sterile conditions to prevent contamination. Water, PEG 300, and NaCl solutions at various concentrations were prepared with fresh DMEM medium. Before applying them on cells, prepared solutions were pre-warmed to 37°C in a water bath to eliminate temperature-induced effects on cell viability and volume. Medium exchange was performed carefully using sterile pipettes.

## Live cell imaging and immunofluorescence

To quantify cellular and nuclear volume changes after osmotic treatment, cells were seeded onto 35-mm confocal dishes (ibidi, Germany) at 20,000 cells per dish. After overnight incubation to ensure complete cell attachment and spreading, cells were stained with Calcein AM (Thermo Fisher Scientific, USA, 1:1000) and Hoechst 33342 (Thermo Fisher, USA, 1:1000) for 30 minutes. After washing with PBS and replacing the medium with L-15 medium (Thermo), cells were subjected to osmotic treatment with water or PEG 300 (MedChemExpress, USA) and immediately imaged using a confocal microscope (LSM 980, Zeiss, Germany). Three-dimensional cell and nuclear images were reconstructed using Z-stack acquisition at 0.2 μm intervals under 405 nm and 488 nm excitation. In addition, 3D fluorescent images were acquired at 1 - 1.5 min intervals. From the acquired 3D image of cells, cellular and nuclear volumes were quantified using Imaris software (Version 7.4.2).

For FUS immunostaining, cells were washed three times with PBS (Thermo Fisher Scientific, USA) to remove residual NaCl. After fixation with 4% paraformaldehyde (Sigma, USA) for 15 minutes, cells were then permeabilized with 0.25% Triton X-100 in PBS (PBST) for 5 minutes and blocked with 1% bovine serum albumin (BSA, Sigma, USA) in 0.1% PBST solution for 45 minutes. The cells were incubated overnight with anti-FUS primary antibody (Proteintech, Germany, Cat No. 68262-1-Ig, 1:200) at 4 °C. Following incubation with Alexa Fluor 488-conjugated secondary antibody (Invitrogen, Cat No. A-11029) and nuclear counterstain for 1 hour, cells were imaged by confocal microscopy.

## Acknowledgements

Z.Q.C. acknowledges the financial support from the National Natural Science Foundation of China (NSFC) and the Research Grants Council (RGC) of the Hong Kong Joint Research Scheme (Project No. N_HKU750/23), HKU seed fund, HKU Outstanding Yong Researcher Award, and the Shenzhen-Hong Kong-Macau Technology Research Programme (Category C project, no. SGDX20230821091501008).

B.G. acknowledges the financial support from Lo Kwee Seong Foundation and Innovation Technology Commission Fund (Health@InnoHK at Center for Translational Stem Cell Biology).

M.S. acknowledges the financial support from JSPS KAKENHI (Grant Number 25K02242), and Takeda Science Foundation.

## Author contributions

Z.C., Y.H. and J.Z. conceived the idea. Z.C. and B.G. supervised the project. Y.H. designed and led the biological experiments. J.Z. performed the measurement and processed the experimental data. H.X. cultured the cells. Y.W. prepare the gold membrane on the substrate. Y.H. prepare the FND samples. J.Z. and Y.H. wrote the manuscript and supplementary information. M.S. discussed the results and commented on the manuscript.

## Competing interests

Z.C., Y.H. and J.Z. are inventors on the NEW US Provisional patent application no. 63/994,457 entitled "Methods and Compositions for Nanodiamond-Based Thermal Profiling of Living Cells". The remaining authors declare no other competing interests.

# Supplementary Information

# All-optical intracellular thermal profiling using nanodiamond-based "thermal radar"

Jiahua Zhang[1,†], Yong Hou[1,†], Xinhao Hu[1], Yicheng Wang[1], Madoka Suzuki[3], Bo Gao[4,5,*], Zhiqin Chu[1,2,*]

[†] Equal contribution
[*] Corresponding authors (Email: bogao@cuhk.edu.hk; zqchu@eee.hku.hk)

[1] Department of Electrical and Electronic Engineering, The University of Hong Kong, Hong Kong, China
[2] School of Biomedical Sciences and School of Biomedical Engineering, The University of Hong Kong, Hong Kong, China
[3] Institute for Protein Research, The University of Osaka, Osaka, Japan
[4] School of Biomedical Sciences, Faculty of Medicine, The Chinese University of Hong Kong, Shatin, Hong Kong, China
[5] Centre for Translational Stem Cell Biology, Tai Po, Hong Kong, China

# Supplementary Note 1. Experimental Setup

The schematic optical path of the home-built upright confocal microscope is illustrated in Figure S1. Fluorescent nanodiamonds (FNDs) are excited using a 532 nm continuous wave (CW) green laser. The emitted fluorescence is directed through a dichroic mirror (Thorlabs, DMLP605R) and long-pass filters (Thorlabs, FELH0600), and subsequently focused onto an avalanche photodiode (APD). Image scanning is performed using a 3-axis piezoelectric stage (PI P-561.3CD).

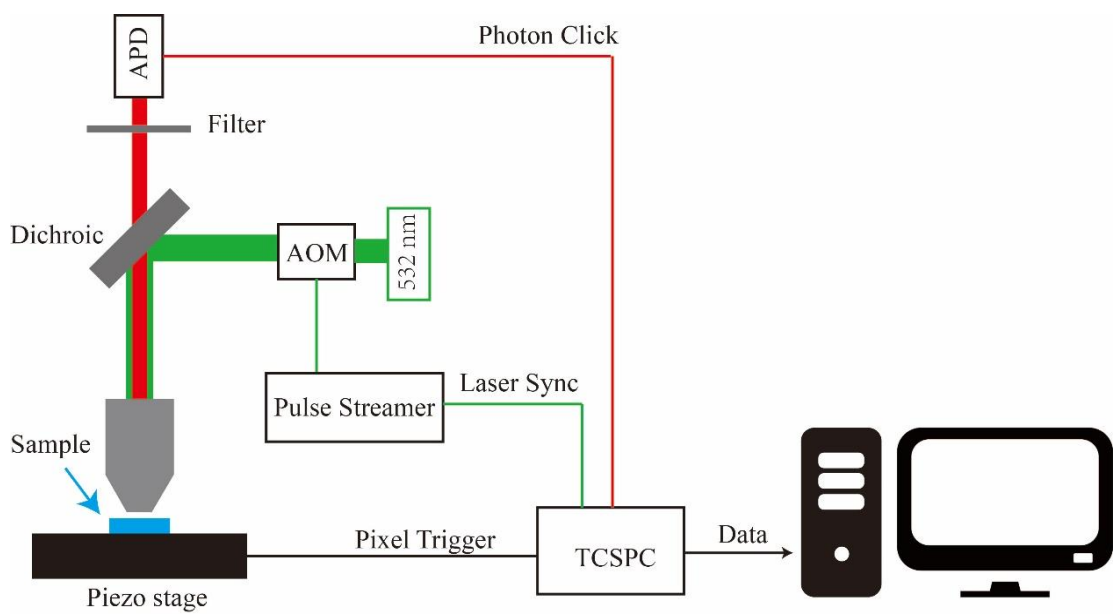

FIG. S1 **Schematic Setup.** Depicts the optical (excitation: green, collection: red) path and signal synchronization setup.

Sinusoidal modulation of the excitation laser is achieved using an acousto-optic modulator (AOM) (Gooch & Housego, AOMO 3250-220 and MHP250-6.6ADM-A1 driver), which controls the laser transmittance in response to an input voltage. A signal generator (Swabian Instruments, Pulse Streamer 8/2) produces a sinusoidal analog voltage signal that drives the AOM, enabling sinusoidal modulation of the laser intensity. The relationship between the input voltage and laser transmittance was characterized, revealing a nonlinear response as shown in Figure S2a. To ensure accurate sinusoidal modulation, only the linear region of the response (0.55–0.75 V) was utilized.

Photon pulses generated by the APD are sent to a time-correlated single photon counting (TCSPC) system (Swabian Instruments, Time Tagger Ultra) to record the fluorescence waveform. Photon counting is synchronized by a TTL signal generated by the Pulse Streamer. Photon histograms are acquired and further processed using Python-based analysis.

The analog sampling rate of the Pulse Streamer is 125 MSa/s, corresponding to an 8 ns time step for the generated sinusoidal waveform. Therefore, the period of the sinusoidal modulation must be an integer multiple of 8 ns. The frequencies used for measurement are: 5, 10, 20, 25, 40, 50, 100, 125, 200, 250, 500, 625, 1000, 1250, 2500, and 5000 kHz. Frequency selection is limited by the capabilities of the Pulse Streamer; use of signal generators with higher analog sampling rates would enable extension to additional frequency points.

## Supplementary Note 2. Frequence Domain Lifetime Measurement

The decay of a single fluorophore from an excited state to the ground state typically obeys an exponential distribution. The decaying intensity over time $I(t)$ will follow equation [1]:

$$I(t) = I_0 e^{-\frac{t}{\tau}} \tag{1}$$

in which $I_0$ is the initial fluorescence at $t = 0$, directly after the excitation pulse, and $\tau$ corresponds to the fluorescence lifetime of that fluorescent species. This equation defines fluorescence lifetime as the time $t$ at which the intensity decreases to 1/e of its initial value $I_0$.

Two approaches were used to measure the fluorescence lifetime: the time-domain and frequency-domain methods. The time domain methods are time-tagging methods that precisely measure photon arrival time relative to the laser pulse. Time-tagging approaches involve assigning a timestamp to each detected photon based on its arrival time after the excitation pulse, allowing for precise measurement of the fluorescence decay kinetics and the determination of fluorescence lifetime. In the time domain methods, the excited laser has an ultrashort pulse duration (~fs), which is assumed as a Dirac delta function compared to the lifetime (~ns). The measured exponential decay is the convolution of the laser and fluorescence exponential distribution. When the laser is considered as a Dirac delta function, the measurement result is the fluorescence dynamic itself, and without further deconvolution process. As we don't have ultrafast laser source for the experiment, we focus on frequency domain implementation methods.

The frequency domain methods are characterized by the usage of a periodic modulated excitation, for which the finite fluorescence lifetime results in a phase delay and demodulates the emission. In theory, the sinusoidal modulated excitation laser $L(t)$ the following equation:

$$L(t) = P + \Delta P \cos(\omega t) \tag{2}$$

So that $\Delta P / P = m_L$ is the modulation amplitude of the incident light. The fluorescence emission is forced to respond with the same frequency, but the phase shift and modulation will be different:

$$I_{FM}(t) = I_0 P \tau + I_0 \Delta P \tau \frac{1}{\sqrt{1+\omega^2 \tau^2}} \cos(\omega t - \varphi_\tau) = A + B \cos(\omega t - \varphi_\tau) \tag{3}$$

where $A = I_{DC} = I_0 P \tau$, and $B = I_0 \Delta P \tau / \sqrt{1 + \omega^2 \tau^2}$. The lifetime can be demodulated from phase or modulated amplitude at given modulation frequency with the following well-known equation [2]:

$$\omega \tau_\varphi = \tan \varphi_\tau \tag{4}$$

$$m_{FM} = \frac{\frac{B}{A}}{m_L} = \frac{1}{\sqrt{1+\omega^2 \tau_m^2}}. \tag{5}$$

For some species, the decaying intensity follows a multi-exponential distribution with the lifetime determined by the weighted contribution ($\alpha_i$) of fluorescence lifetimes ($\tau_i$) [1]:

$$I(t) = I_0 \sum_i \alpha_i e^{-\frac{t}{\tau_i}} \tag{6}$$

For example, the fluorescence of NV centers exhibits a biexponential decay process, expressed as $I_0(ae^{-t/\tau_1} + be^{-t/\tau_2})$ with $a + b = 1$. For such conditions, the amplitude modulation or phase shift response can be similarly defined as follows [2]:

$$\tan \varphi_\tau = \frac{N_\omega}{D_\omega} \tag{7}$$

$$m_{FM} = \sqrt{N_\omega^2 + D_\omega^2} \tag{8}$$

where $N_\omega = \frac{\frac{a\omega\tau_1^2}{1+\omega^2\tau_1^2}+\frac{b\omega\tau_2^2}{1+\omega^2\tau_2^2}}{a\tau_1+b\tau_2}$, and $D_\omega = \frac{\frac{a\tau_1}{1+\omega^2\tau_1^2}+\frac{b\tau_2}{1+\omega^2\tau_2^2}}{a\tau_1+b\tau_2}$.

Compared to using one frequency, normally, the fluorescence lifetime is determined by fitting the modulated amplitude or phase change response over a wide frequency range. In our work, we prefer the modulated amplitude response curve to fit the lifetime.

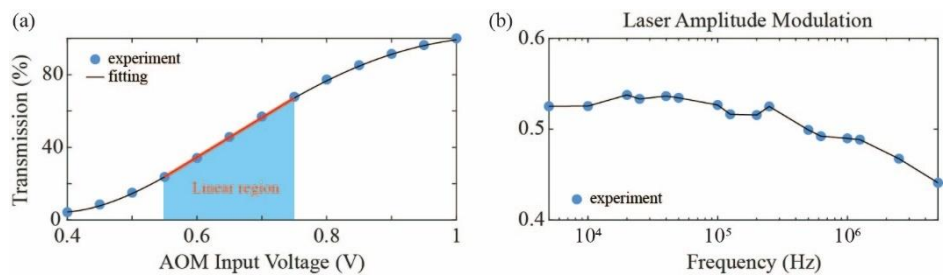

FIG. S2 **Laser property for experiment**. (a) Relationship for laser transmission and AOM input voltage. Voltage in linear region is used for generating sinusoidal modulation. (b) The modulated amplitude of laser over the frequency band from 5 kHz to 5 MHz

First, we measured the laser modulation response, since the system response varies across frequencies due to the optical components and the AOM. To do this, we removed the long-pass filter and measured the reflected laser signal to obtain the laser modulation curve (Fig. S2b). As shown in Fig. S3a, the saturated curve of one FND is measured. To decrease the nonlinear effect, we control the laser power is around 100 μW which is smaller than saturated power (~370 μW). We then recorded the fluorescence waveform at different modulation frequencies and computed the modulation amplitude by dividing the fluorescence amplitude by the laser amplitude (Fig. S3b). At low frequencies, the modulation amplitude is approximately 0.9 rather than unity. There are several reasons leading to this phenomenon, most importantly related to quantum yield. Therefore, we further normalize the modulation curve by one coefficient to fitting the lifetime:

$$m_{measured\ FM} = \eta \cdot m_{FM} \tag{9}$$

It is worth noting that nanodiamonds with higher nitrogen concentrations have higher fluorescence intensity when excited at the same laser power, which shortens the excited-state lifetime and thus reduces fluorescence quantum yield through non-radiative decay [3]. In cellular measurements, higher fluorescence can reduce measurement time and improve the temporal resolution of dynamic measurements, although it comes with a lower quantum yield and coefficient of approximately 0.7-

0.9.

We performed a statistical analysis of the lifetimes of FNDs and compared them with those of bulk diamond (Fig. S3c). The long-lifetime component in FNDs is 19.1 ± 3.07 ns, slightly larger than that in bulk diamond (15.7 ± 0.45 ns). The uncertainty is also greater for FNDs, reflecting their heterogeneity. This value will be used for subsequent thermal conductivity fitting.

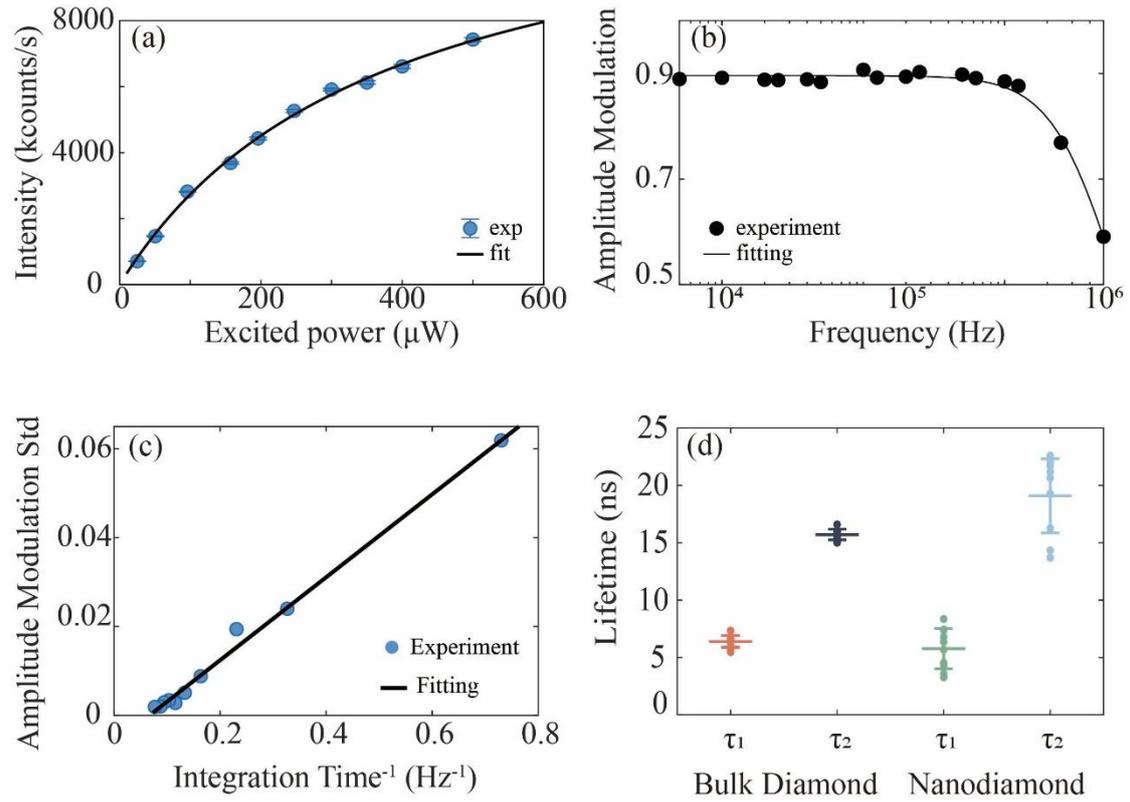

FIG. S3 **Fluorescence property**. (a) The saturation curve of one FND. The saturation power is around 370 μW. (b) The modulated amplitude of fluorescence which has been divided by the laser modulated amplitude over the frequency band from 5 kHz to 5 MHz. For the curve fitting, it should be further normalized by a coefficient. (c) The relationship between standard deviation and integration time for one frequency for FND with 1 Mcounts/s. (d) Measured lifetime. The data is compared between bulk diamond and FNDs. The statistical number of FNDs and position in bulk diamond are n = 10, respectively.

## Supplementary Note 3. Thermal Conductivity Measurement

Our method is an improvement on FDTR, retaining the most part theory of the heating pump laser. A brief overview of the heating theory is as follows. The intensity of the heating laser in real space and time domain is expressed as [4]:

$$p(r,t) = \frac{2A_1}{\pi w_1^2} \exp\left(-\frac{2r^2}{w_1^2}\right) e^{i\omega_0 t} \tag{10}$$

It is continuous wave modulated by a sinusoidal function at frequency $\omega_0$. $A_1$ is the average power of the heating laser, which has a Gaussian distribution in space with a $1/e^2$ radius of $w_1$. The frequency domain expression of the heating laser intensity can be obtained by the Hankel transform on space and the Fourier transform on time [4]:

$$P(k,\omega) = A_1 \exp\left(-\frac{\pi^2 w_1^2 k^2}{2}\right) 2\pi\delta(\omega - \omega_0) \tag{11}$$

The heating surface temperature response in the frequency domain is the product of the heat input P and the thermal response function of the system $\hat{G}$ [4]:

$$\Theta(k,\omega) = P(k,\omega)\hat{G}(k,\omega) \tag{12}$$

The thermal response function of the multilayer stack system for solid material, such as cover glass, sapphire wafer, and silicon wafer, have been studied for many years. The details can be found in references [4]. For water and other liquid phase samples, a bidirectional heat conduction thermal model is employed [5].

Inverse Hankel transform on the equation above gives the temperature distribution on the surface as a result of modulated heating:

$$\Theta(r,\omega) = \int_0^\infty P(k,\omega)\hat{G}(k,\omega) J_0(2\pi k r) 2\pi k dk \tag{13}$$

The inverse Fourier transform of $\Theta(r,\omega)$ gives the surface temperature response $\theta(r,t)$. As NV center is an atomic sensor, the temperature detected can be assumed as the one at center position. The temperature response in the frequency domain can be simply as:

$$\Delta T(\omega) = \Theta(0,\omega) = \int_0^\infty P(k,\omega)\hat{G}(k,\omega) 2\pi k dk$$

$$= A_1 \int_0^\infty \hat{G}(k,\omega) \exp\left(-\frac{\pi^2 w_1^2 k^2}{2}\right) 2\pi k dk \, 2\pi\delta(\omega - \omega_0) \tag{14}$$

Thus, the surface temporal temperature response detected by NV centers

$$\Delta T(t) = A_1 e^{i\omega_0 t} \int_0^\infty \hat{G}(k,\omega_0) \exp\left(-\frac{\pi^2 w_1^2 k^2}{2}\right) 2\pi k dk \tag{15}$$

It can be further written in the real space as a sinusoidal oscillates frequency $\omega_0$:

$$T(\omega_0, A_1) = Z(\omega_0, A_1) \cos(\omega_0 t + \varphi_T) \tag{16}$$

Where $\tan\varphi_T = \frac{Imag(Z(\omega_0, A_1))}{Real(Z(\omega_0, A_1))}$, and $Z(\omega_0, A_1) = A_1 \int_0^\infty \hat{G}(k,\omega_0) \exp\left(-\frac{\pi^2 w_1^2 k^2}{2}\right) 2\pi k dk$.

Actually, the sinusoidal modulated excitation laser $L(t)$ [Eq. (1)] has a DC component, the finial temperature response is:

$$T(t) = Z(0,P) + Z(\omega, \Delta P) \cos(\omega t + \varphi_T) \tag{17}$$

The theory above illustrates the laser-heating mechanism, and the next shows how fluorescence thermometry reports temperature. We use fluorescence intensity as the physical observable to quantify temperature. The dependence of fluorescence intensity on temperature is given by:

$$I(t,T) = I(t)(1 - \alpha T) \tag{18}$$

The sinusoidal modulated green laser [Eq. (1)] not only excites the fluorescence of NV centers but also induces temperature oscillations. The finally temporal fluorescence intensity is combined with the [Eq. (3)], [Eq. (17)], and [Eq. (18)] as:

$$I(t,T) = I_{FM}(t)\big(1 - \alpha T(t)\big) \tag{19}$$

After simplification, the time-domain fluorescence intensity comprises three components: a DC term, a component at frequency ω, and a component at 2ω. Because the 2ω term has a smaller amplitude than the ω term, the waveform remains approximately sinusoidally modulated. We therefore focus on the DC and ω components. The modulation amplitude of fluorescence signal is defined as:

$$m_{FM,\omega} = \frac{A_{FM,\omega}}{DC_{FM,\omega}} \tag{20}$$

The DC term is defined as:

$$DC_{f,\omega} = I_{DC}\big(1 - \alpha Z(0,P)\big) - I_{DC}\frac{\Delta P m_{FM} \alpha Z(\omega, \Delta P)}{2P}\cos(\varphi_\tau + \varphi_T) \tag{21}$$

The component at frequency ω is defined as:

$$A_{FM,\omega} = \sqrt{\left\{I_{DC}\frac{\Delta P m_{FM}}{P}\left[1 - \alpha Z(0,P) - \alpha\frac{P}{\Delta P m_{FM}}Z(\omega, \Delta P)\cos(\varphi_\tau + \varphi_T)\right]\right\}^2 + [\alpha I_{DC} Z(\omega, \Delta P)\sin(\varphi_\tau + \varphi_T)]^2} \tag{22}$$

Finally, the function to fit the experiment data is similar to [Eq. (9)], defined as:

$$m_{measured\ FM,\omega} = \eta \cdot \frac{m_{FM,\omega}}{m_L} \tag{23}$$

The Fig. S4 shows the amplitude experiment data, and the parameters used in the fits for cover glass, water as the samples using two different heat conduction models. In Ref. [6], when the pillar radius is at least five times larger than the heating laser spot size, the thermal response is well described by a semi-infinite model. Because the cell size (~100 μm) greatly exceeds both the heating laser spot size (~λ) and the sensor size (~100 nm), we model the cell as a single layer. The intracellular thermal conductivity is then obtained by fitting the FND fluorescence measured in cells with a bidirectional heat-conduction model. Our measurements are consistent with those of Ref. [7], validating the applicability of the multilayer model to biological systems. We also measured the in-situ cell height and incorporated it into the fitting of the amplitude-modulation data.

It should be noted that radial heat transfer dominates the thermal dissipation in the measurement. Consequently, the measured thermal conductivity is the geometric average thermal conductivity $\sqrt{\kappa_r \kappa_z}$ for anisotropic materials. In our sample set, water, cover glass, and silicon are isotropic, whereas sapphire is anisotropic. For sapphire, the thermal conductivities parallel and perpendicular to the crystal axis are $\kappa_\parallel = 34\ \text{W} \cdot \text{m}^{-1} \cdot \text{K}^{-1}$ [8] and $\kappa_\perp = 25\ \text{W} \cdot \text{m}^{-1} \cdot \text{K}^{-1}$, respectively, yielding a geometric mean of approximately 29.5 W·m⁻¹·K⁻¹, consistent with our measurement.

In addition, there is one important parameter should be discussed, called "sensitivity", defined to characterize the dependence of signal on different parameters (interface thermal conductance, heat capacity, and thermal conductivity, etc.) across the large frequency range. Such "sensitivity" is different with the one used in the manuscript to characterize the minimum distinguishable quantity of thermal conductivity. Such sensitivity can help us choose the frequency range to highly efficient

analysis the desired parameter. In our method, it is analyzed following established protocols [9]:

$$S_\kappa = \frac{\partial \ln m_\omega}{\partial \ln \kappa} \qquad (24)$$

Where $S_\kappa$ quantifies the percentage change in modulation amplitude per percentage change in thermal conductivity κ. For example, a sensitivity value of 0.4 means that there would be 0.4% change in modulation amplitude if the thermal conductivity is changed by 1%. As shown in the Figure S4b, sensitivity is higher for high-κ materials and at low frequencies, due to greater temperature changes amplifying amplitude reductions.

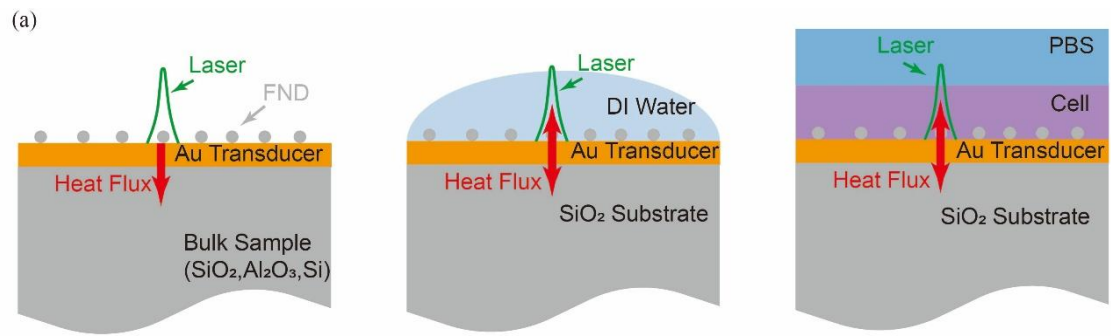

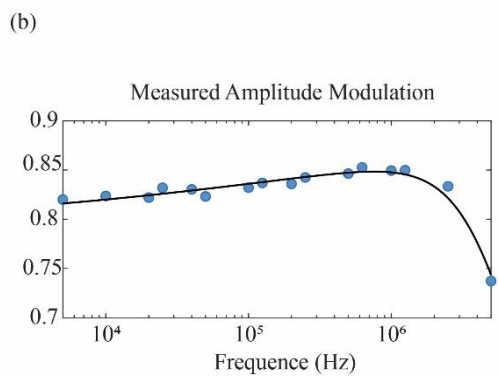

| Sample: c-SiO₂, T = 300 K, Heating Power = 119 + 64.3·sin(ωt) μW | |
|---|---|
| $\kappa$, (W/m·K) | 1.416 |
| $G$, (MW/m²·K) | 84.3 |
| $\tau_1$, (ns) | 3.7 |
| $\tau_2$, (ns) | 24.3 |
| Weight for Lifetime 1 | 0.879 |
| coefficeient $\alpha$, (%/K) | 1.78 |
| coefficeient $\eta$ | 0.934 |

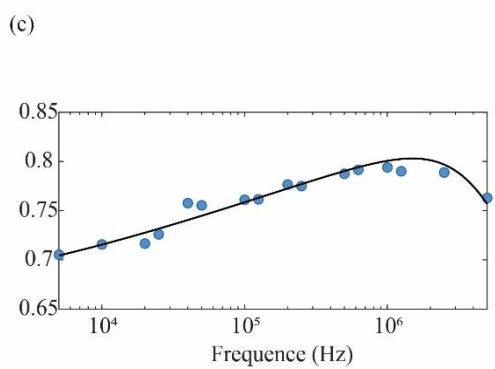

| Sample: H₂O, T = 300 K, Heating Power = 119 + 64.3·sin(ωt) μW | |
|---|---|
| $\kappa$, (W/m·K) | 0.59 |
| $G$, (MW/m²·K) | 43.9 |
| $\tau_1$, (ns) | 6.62 |
| $\tau_2$, (ns) | 25.1 |
| Weight for Lifetime 1 | 0.931 |
| coefficeient $\alpha$, (%/K) | 3.53 |
| coefficeient $\eta$ | 0.9 |

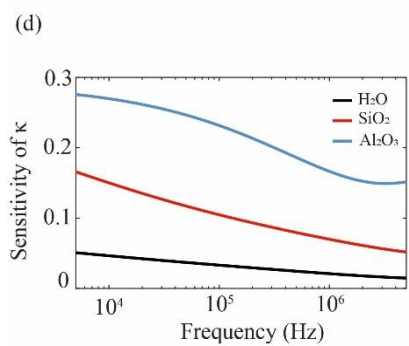

FIG. S4 **Thermal conductivity measurement supplement data.** (a) Thermal conductivity measurement illustration for different samples. (b) Amplitude experiment data, and the parameters used in the fits for cover glass. It is chosen as example for bulk samples (c) Amplitude experiment data, and the parameters used in the fits for water. It is chosen as example for aqueous media. (d) Sensitivity of the thermal model to thermal conductivity changes.

# Supplementary Note 4. Heat Transfer Simulation

We developed a simulation to investigate the thermal penetration depth in the liquid phase, which governs our ability to measure subcellular organelles located above the heating interface. For efficiency, the geometry was reduced from a full 3D model to a 2D axisymmetric model (Figure S5a). The domain comprises two regions: a cover glass substrate (600 μm length × 150 μm height) and an overlying water layer (600 μm length × 500 μm height). These dimensions were chosen so that halving the domain size changes the temperature at the heating center by less than 0.01 K, ensuring size-invariant results. Because the heating spot size is comparable to the laser wavelength, the water and glass thickness are sufficient to approximate semi-infinite media. Thus, all external boundaries are fixed at room temperature in the simulation because the experiments are conducted at room temperature. The glass–water interface is modeled with a thin-film boundary condition representing a 50 nm-thick gold membrane, and a boundary heat source is applied at this interface to simulate the laser-induced heat flux. The whole model is meshed with triangle mesh for simulation (Figure S5b).

For steady state simulation, the heat source used in the simulation is defined as:

$$Q_0 = \frac{2A_1}{\pi w_1} \exp\left(-\frac{2r^2}{w_1^2}\right) \tag{24}$$

Where the heating laser power $A_1$ is set 100 μW, $w_1 = 0.61\lambda/NA$ in the simulation. In the liquid-phase experiment, we used a water-dipping objective (Nikon N40X-NIR, NA = 0.80). The simulation predicted a temperature rise of approximately 6.96 K. To compare with the simulation, we employed NV-center-based all-optical thermometry [10] to measure the temperature increase; the calibration of ZPL position and measurement results are shown in Figure S6c. By comparing spectra acquired at laser powers of 150 μW and 50 μW, we determined that a 100 μW increase in incident laser power raises the temperature by about 3.89 K. From the experiment result, the laser absorption efficiency for heating can be estimated to be 55.89%.

For transient state simulation, the heat source used in the simulation is defined as:

$$Q_0 = \frac{2A_1}{\pi w_1} \exp\left(-\frac{2r^2}{w_1^2}\right)(1 + 0.54 \cdot \cos(\omega t)) \tag{25}$$

We performed simulations at frequencies of 5, 10, 20, 50, 80, 100, 200, 500, 800, 1000, 2000, and 5000 kHz. For each frequency, we extracted the TPD from the curve in Figure 2e to obtain the TPD-versus-frequency curve shown in Figure 2f.

As the cell is heterogeneity, it is hard to simulate the real condition. We can simply estimate the temperature rise caused by laser heating in the presence of cells. Using the measured mean thermal conductivity in Figure 3c, we assume a uniform body for the cell in this simulation. The 3D model, which is shown in Figure S5c, is used in the simulation. The domain comprises three regions: a cover glass substrate (600 μm length × 600 μm weight ×150 μm height), an overlying PBS layer (600 μm length × 600 μm weight × 500 μm height), and the cell model (40 μm radius × 10 μm height). The boundary conditions are set same with the liquid simulation. For transient state simulation, the heat source used in the simulation is defined as:

$$Q_0 = \frac{2A_1}{\pi w_1} \exp\left(-\frac{2[(x-x_0)^2 + (y-y_0)^2]}{w_1^2}\right) \tag{26}$$

Where $(x_0, y_0)$ is the heat center to simulate different positions in the cells. Under the nucleus, we set $(x_0, y_0) = 0$; under cytoplasm, we set $(x_0, y_0) = (20\ \mu m, 0)$. Considering the laser absorption efficiency of 55.89% estimated from the preceding liquid-phase experiment, an effective heating power of approximately 67 µW (incident laser power of 120 µW × 55.89%) was used in this simulation. Under this condition, the simulation predicted a temperature rise of approximately 5.52 K under the cytoplasm and 4.69 K under the nucleus. This difference reflects the different local geometry and heat dissipation paths between the cytoplasmic and nuclear regions.

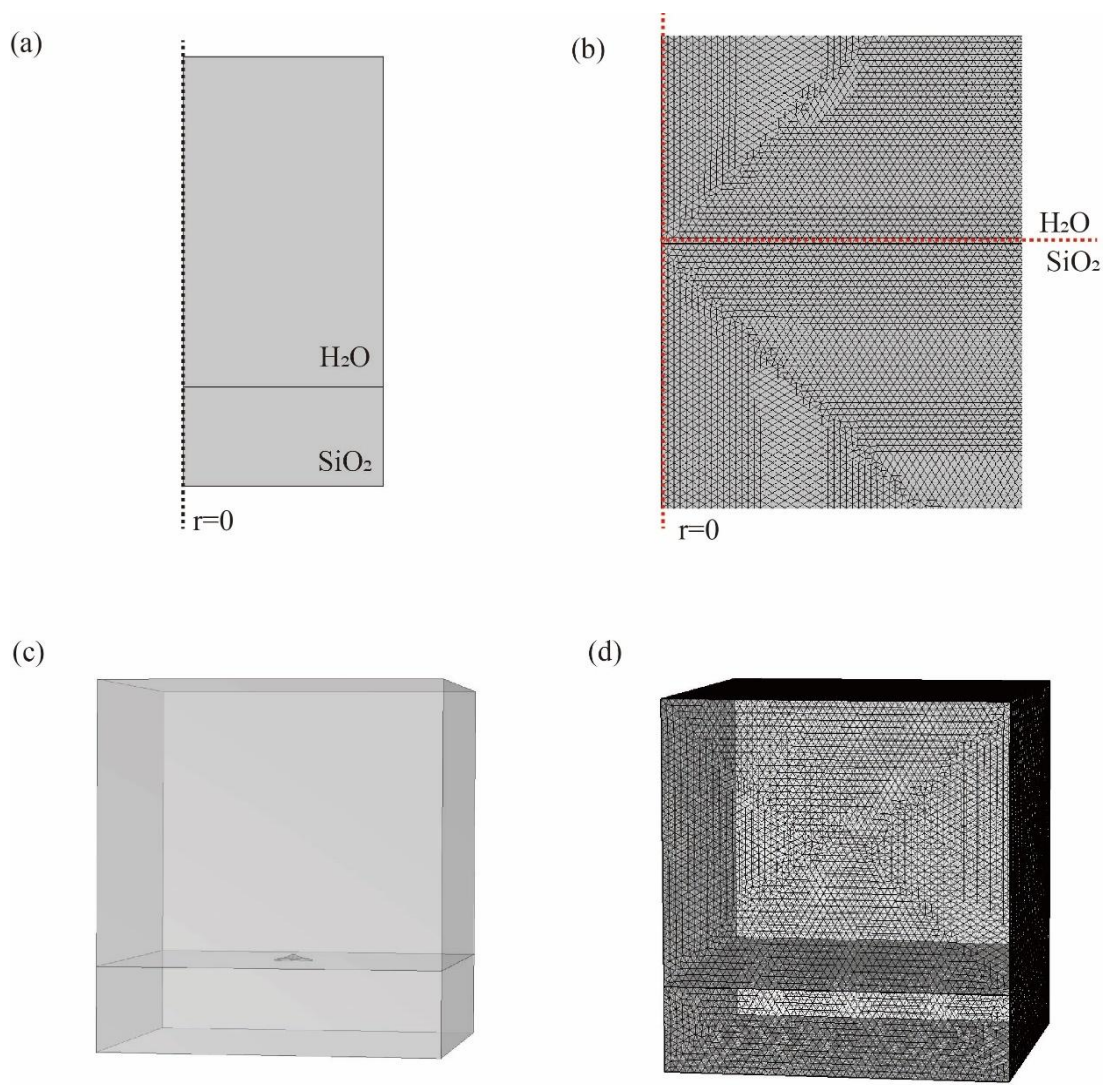

Fig S5 **Simulation model and mesh.** (a) liquid phase simulation model. Because the heating source is at the center, 2D axisymmetric model is used for this kind of simulation. (b) corresponding mesh for model in (a). (c) cell simulation model. (d) corresponding mesh for model in (c).

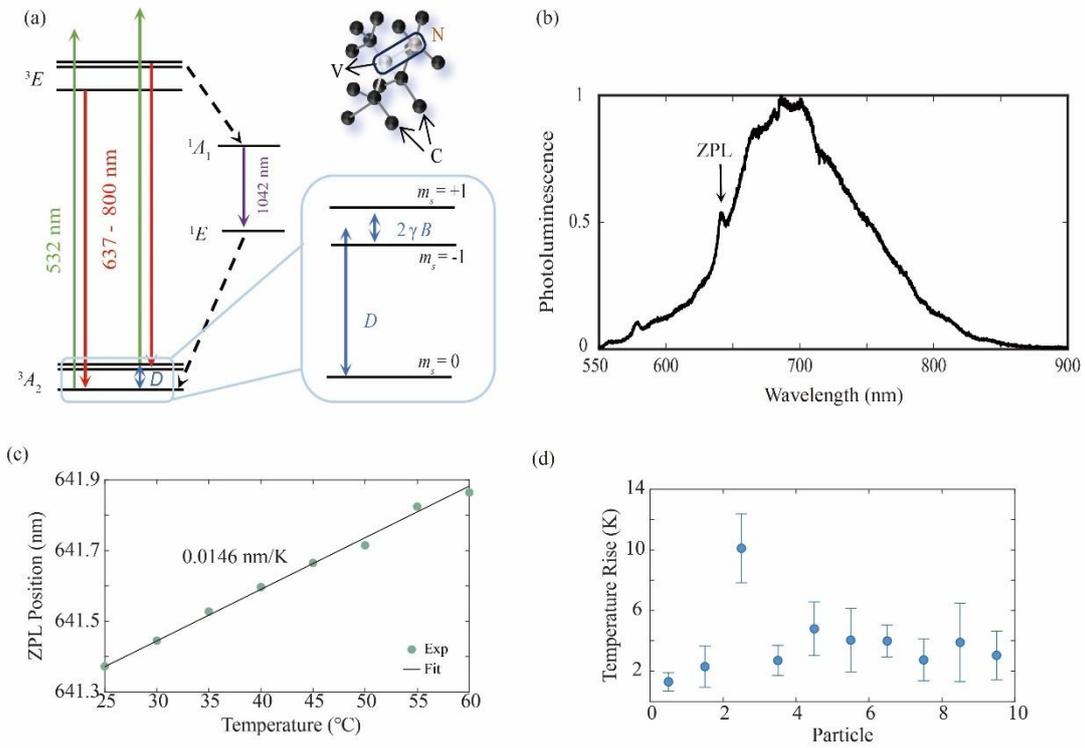

Fig S6 **all-optical thermometry based on NV spectrum.** (a) energy level of NV center. (b) NV fluorescence spectrum. (c) The relationship between temperature and ZPL position. (d) number of 10 FNDs measured the temperature rise when a 100 μW laser heat the gold membrane.

# Supplementary Note 5. Phototoxicity Analysis

To evaluate the phototoxicity of laser, Hela cells were seeded into a cover glass with an imprinted 500 um cell location grid (ibidi, Grid 500) at a density of 4-6 per grid. After 24 hours, the cells were illuminated with 120 µW green laser under confocal microscopy (0.95 NA,). Such laser power is chosen according to the heating power in measuring thermal conductivity experiment. Over a five-day period, we evaluated cellular viability by measuring the proliferation rate. The result is shown in Figure S7. The proliferation rate is similar to the one before and after laser shining. Such laser power in the confocal has less effect phototoxicity for the cell.

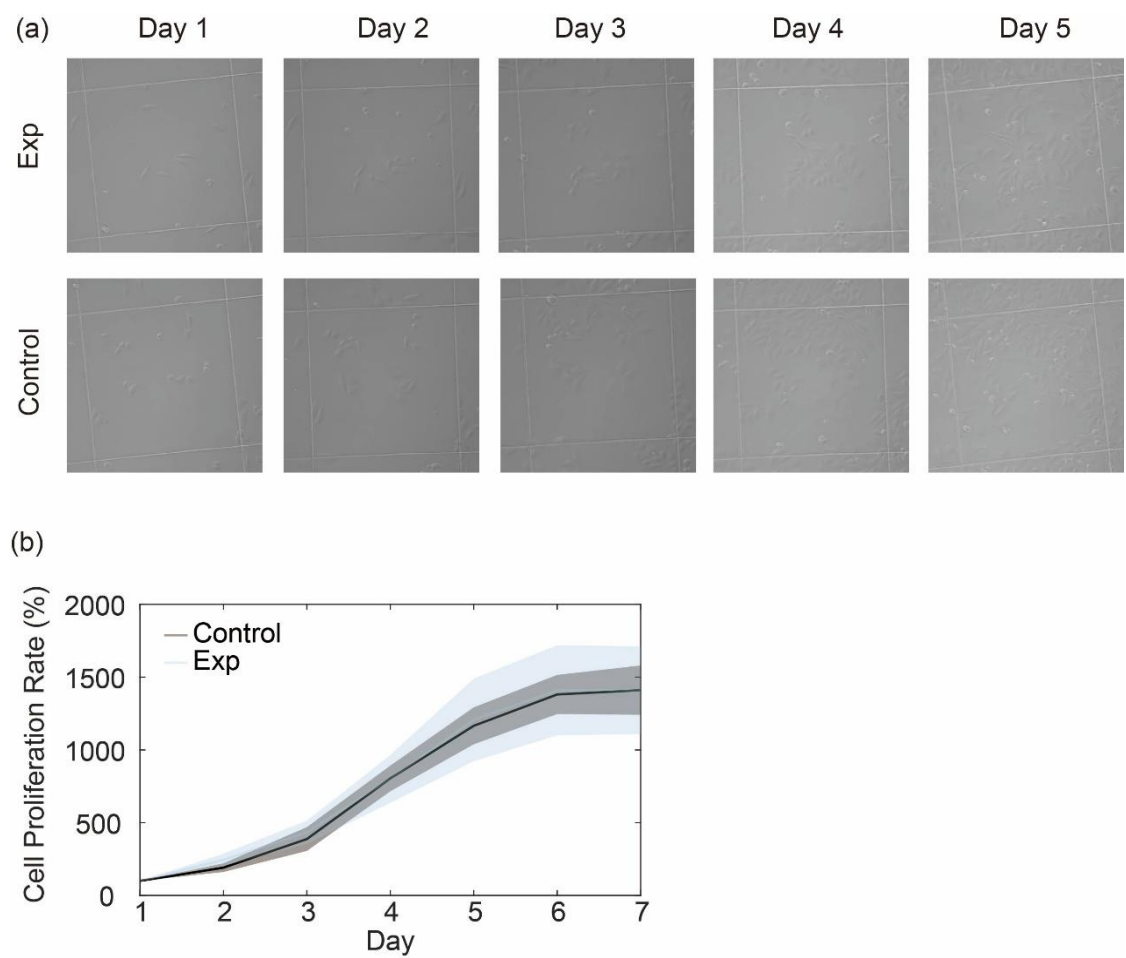

Fig S7 Phototoxicity Analysis (a) images of cell for 5 days (b) The proliferation rate curve over the 5 days.

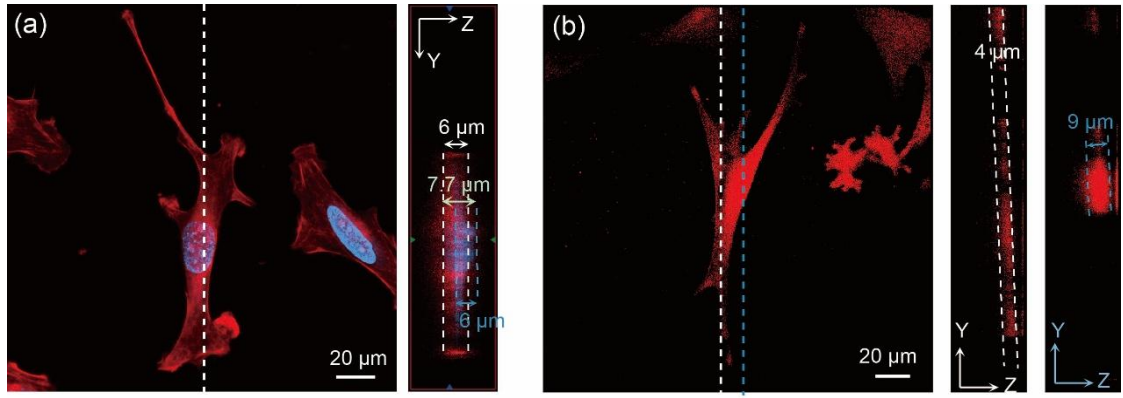

Fig S8 **Distance about the substrate and nucleus measured by three-dimension fluorescence images**. (a) Confocal xy and yz image of a fixed Hela cell (nucleus height: 6 μm; subnuclear bottom to cell bottom: 1.7 μm; cytoplasm height: ~ 6 μm). (b) Confocal xy and yz image of a fixed 3T3 cell (blue line: nucleus height: 9 μm; white line: cytoplasm height: ~ 4 μm).

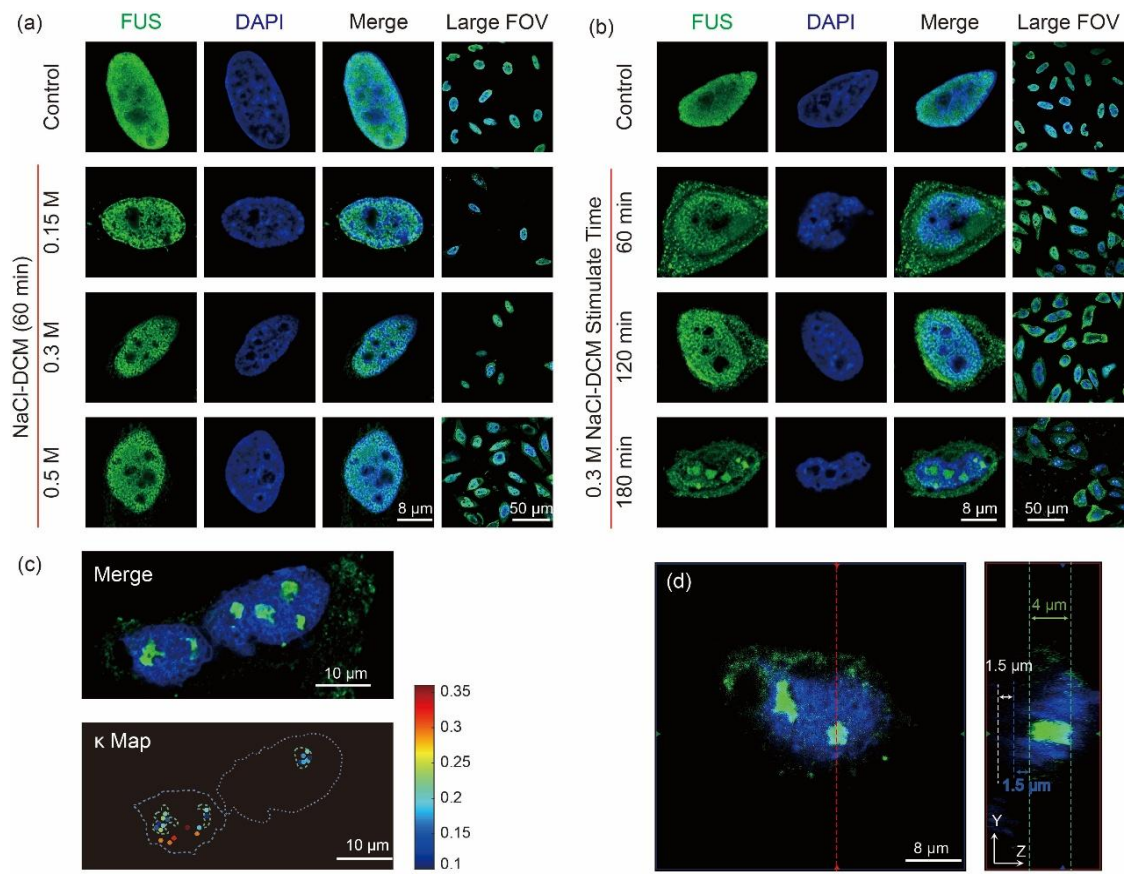

Fig S9 **Measurement of thermal conductivity in liquid-liquid phase separation**. (a) Confocal fluorescence image of FUS aggregation by 0.15 M, 0.3 M and 0.5 M NaCl dissolved in the PBS for 60 min. The image of FUS in the complete medium without NaCl as control experiment. (b) Confocal fluorescence image of FUS aggregation by 0.3 M NaCl dissolved in the PBS for 60, 120 and 180 min. The fluorescence image of FUS in the complete medium without NaCl as control experiment. (c) Confocal xy fluorescence image and thermal conductivity map of a fixed HeLa cell (180 min, 0.3 M NaCl). (d) Confocal xy and yz image of a fixed HeLa cell (FUS aggregation height: 4μm; subnuclear bottom to FUS bottom: 1.5 μm; subnuclear bottom to cell bottom: 1.5 μm).

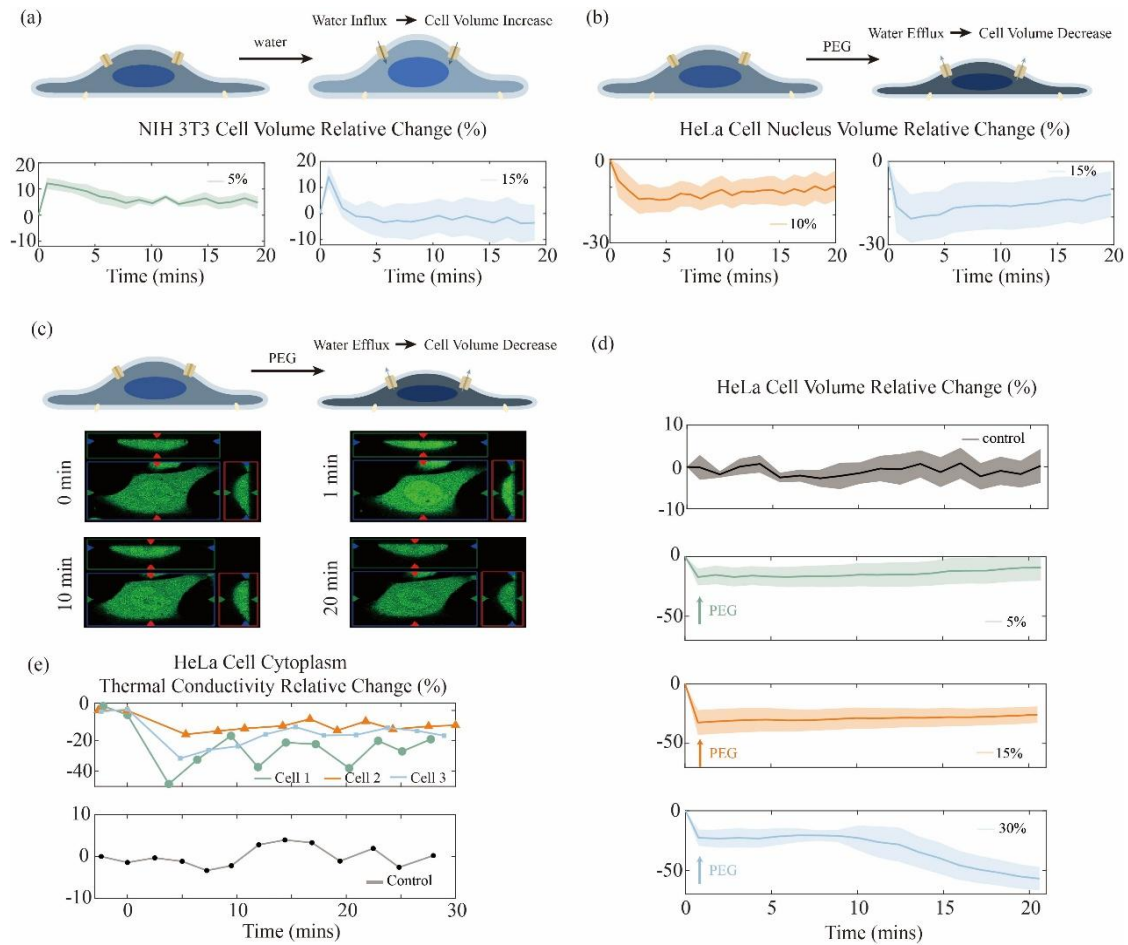

Fig S10 **Measurement of the change of thermal conductivity following osmatic stress in a single living cell at 33°C**. (a) The cell volume relative change curve of NIH 3T3 cells after adding water solutions with volume fractions of 5% and 15%. (b) The cell nucleus volume relative change curve of HeLa cells after adding peg solutions with volume fractions of 10% and 15%. (c) The scheme of cells under osmatic compressions, which the cells volume will decrease. Four fluorescence orthogonal projections of one of the HeLa cells after adding peg solution with 5% volume fractions at 0 min, 1min, 10min, and 20 min are listed below the illustration scheme. (d) The cell volume relative changes of HeLa cells after adding PEG solutions with different volume fractions. We add the volume fractions are 0%, 5%, 15% and 30% respectively. (e) The measurement of thermal conductivity relative change after adding peg solution with 7.5% volume fractions.

# Supplementary Note 6. The effect of cell motion on the thermal conductivity measurement

In the absence of pharmacological perturbation, confocal fluorescence imaging shows that single cells migrate at approximately 1 μm per 40 minutes; under these conditions, single-point thermal conductivity tracking exhibits only 1.5–2% fluctuations, indicating that such small displacements do not measurably bias the dynamics via spatial heterogeneity. The standard deviation of a single-point measurement is primarily determined by FND fluorescence intensity. Over extended observation windows (>2 h), however, cell spatial heterogeneity can introduce substantially larger oscillations; maintaining an in-situ sensor at the region of interest may require switching to a different FND, but quantitative comparisons across distinct FNDs are discouraged because particle heterogeneity can contribute approximately 10% variability.

Under hypotonic stress (water addition), nuclear motion accompanying cell swelling is evident (Figure S11a). Because FNDs are stochastically distributed, an off-center nuclear FND may drift outside the nucleus, compromising the in-situ nature of the measurement and confounding the dynamics with nuclear spatial heterogeneity; unlike a fluorescence-volume trajectory, a stable time-resolved nuclear thermal conductivity trace is therefore not feasible, and we did not pursue it in hypotonic conditions.

In a PEG addition experiment, the tracked cell underwent apoptosis (Figure S11b) and the FND traversed the nuclear envelope, producing a marked change in apparent thermal conductivity (Figure S11c). The transition indicates higher conductivity at the nuclear boundary relative to the nucleoplasm, consistent with Ref. [7]. This case study highlights how spatial heterogeneity and sensor motion can dominate the measured dynamics, in contrast to the stable behavior observed in Figure 5e.

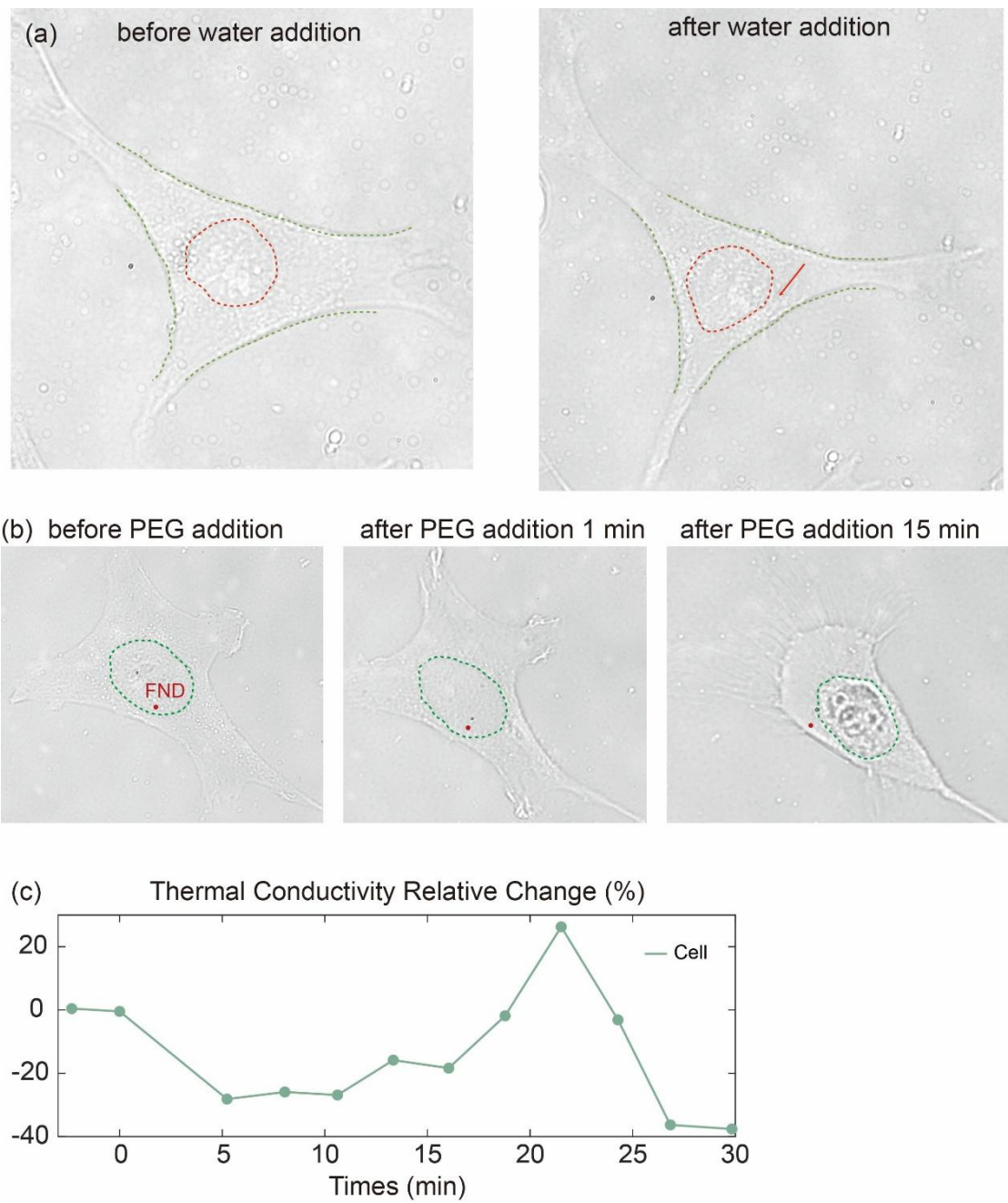

Fig S11 **Cell motion influence** (a) nucleus motion before and after water addition in NIH 3T3 cells. Nucleus is surrounded by dotted curve. The line on the right indicates the direction of cell nucleus movement. (b) The measured FND moves outside the nucleus during cell apoptosis after PEG addition. (c) Corresponding thermal conductivity relative change when the FND moves outside the nucleus in (b).